\documentstyle[aas2pp4]{article}
\lefthead{Fassnacht et al.}
\righthead{I. A Determination of H$_0$ with the CLASS
	   Gravitational Lens B1608+656}
\begin{document}

\slugcomment{Submitted to \apj}
\title{A Determination of $H_0$ with the CLASS Gravitational Lens B1608+656:
I.~Time Delay Measurements with the VLA\footnote{The National Radio
Astronomy Observatory is operated by Associated Universities, Inc.,
under cooperative agreement with the National Science Foundation.}}

\author{
	   C.\ D.\ Fassnacht\altaffilmark{2,3},
	   T.\ J.\ Pearson\altaffilmark{2},
	   A.\ C.\ S.\ Readhead\altaffilmark{2},
	   I.\ W.\ A.\ Browne\altaffilmark{4},
	   L.\ V.\ E.\ Koopmans\altaffilmark{5},
	   S.\ T.\ Myers\altaffilmark{6},
	   P.\ N.\ Wilkinson\altaffilmark{4}
}
\altaffiltext{2}
{
	   Owens Valley Radio Observatory, 
	   California Institute of Technology, 105-24, 
	   Pasadena, CA 91125.
	   E-mail: cdf@astro.caltech.edu, tjp@astro.caltech.edu, acr@astro.caltech.edu
}
\altaffiltext{3}
{
	   Current Address:  
	   NRAO,
	   P.O.\ Box O,
	   Socorro, NM 87801.
	   E-mail: cfassnac@aoc.nrao.edu
}
\altaffiltext{4}
{
	   NRAL Jodrell Bank,
	   University of Manchester, 
	   Macclesfield, 
	   Cheshire SK11 9DL, UK.
	   E-mail: iwb@jb.man.ac.uk, pnw@jb.man.ac.uk
}
\altaffiltext{5}
{
	   Kapteyn Astronomical Institute,
	   Postbus 800,
	   9700 AV Groningen,
	   The Netherlands.
	   E-mail: leon@astro.rug.nl
}
\altaffiltext{6}
{
	   Alfred P. Sloan Fellow.
	   Department of Physics and Astronomy, 
	   University of Pennsylvania, 
	   209 S. 33rd St., 
	   Philadelphia, PA 19104-6396.
	   E-mail: myers@upenn5.hep.upenn.edu
}

\begin{abstract}

We present the results of a program to monitor the four-image
gravitational lens B1608+656 with the VLA.  The system was observed
over a seven month period from 1996 October to 1997 May.  The 64
epochs of observation have an average spacing of 3.6~d.  The light
curves of the four images of the background source show that the flux
density of the background source has varied at the $\sim$5\% level.
We measure time delays in the system based on common features that are
seen in all four light curves.  The three independent time delays in
the system are found to be $\Delta t_{BA} = 31\pm 7$~d, $\Delta t_{BC}
= 36\pm 7$~d, and $\Delta t_{BD} = 76^{+9}_{-10}$~d at 95\%
confidence.  The uncertainties on the time delays are determined by
Monte Carlo simulations which use fake light curves that have the
characteristics of the observed light curves.  This is the first
gravitational lens system for which three independent time delays have
been measured.  A companion paper presents a mass model for the
lensing galaxy which correctly reproduces the observed image
positions, flux density ratios, and time delay ratios.  The last
condition is crucial for determining $H_0$ with a four-image lens.  We
combine the time delays with the model to obtain a value for the
Hubble constant of $H_0 = 59^{+8}_{-7}$~km~s$^{-1}$~Mpc$^{-1}$ at
95\% confidence (statistical) for $(\Omega_M, \Omega_{\Lambda}) = (1, 0)$.
In addition, there is an estimated systematic uncertainty of
$\pm15$~km~s$^{-1}$~Mpc$^{-1}$ from uncertainties in modeling the
radial mass profiles of the lensing galaxies. The value of $H_0$
presented in this paper is comparable to recent measurements of $H_0$
from the gravitational lenses 0957+561, PG~1115+080, B0218+357, and
PKS~1830$-$211.

\end{abstract}


\section{Introduction}

Even before the discovery of the first gravitational lens system, a
technique for using gravitational lenses to measure the distance scale
of the universe had been developed (\cite{6_refsdal}).  The technique
requires a lens system in which multiple images of the background
source are formed.  A ``map'' of the geodesics along which the light
travels to form the images is constructed and used to predict the
differences in light travel times along the geodesics.  If the
background source is variable, these time delays can be measured as
each image varies in turn.  The ratios between the observed and
predicted delays give the Hubble constant in the assumed world model
$(\Omega_M, \Omega_\Lambda)$.  The use of gravitational lenses for
determining $H_0$ has major advantages over traditional ``distance
ladder'' approaches.  First, the technique gives a direct estimate of
$H_0$ at cosmological distances, where the effects of peculiar
velocities are minimal.  Second, this measurement of $H_0$ is obtained
in one step, without the propagation of errors inherent in the
distance ladder approach.

With the discovery that the ``double quasar'' 0957+561 A,B was a lens
system (\cite{6_0957discovery}), the effort to use lenses to measure
$H_0$ began in earnest.  For many years the effort was hindered by
both the paucity of known lens systems and the difficulty in measuring
time delays in the systems.  In fact, an unambiguous time delay has
only recently been measured in 0957+561 in spite of over ten years of
intensive monitoring (\cite{6_k0957_1}; \cite{6_k0957_2}; \cite{6_o0957};
\cite{6_dh0957}).  Another of the earliest known lenses, PG~1115+080
(\cite{6_1115discovery}), has also just produced measurable time delays
(\cite{6_s1115}).  However, we may have entered a new era for
time delay measurements from gravitational lenses.  One reason for
this is the accelerated rate of discovery of new lenses from
systematic radio surveys.  The Jodrell-VLA Astrometric Survey (JVAS;
\cite{6_jvas1}; \cite{6_jvas2}; \cite{6_jvas3}) has produced 6 new lenses, 
and time delays have been measured for one of them (B0218\+357;
\cite{6_b0218}).  The ongoing Cosmic Lens All-Sky Survey (CLASS;
\cite{6_class}) has found 12 new lenses since it began in 1994.  This
paper reports the first measurement of time delays from a CLASS lens.

The CLASS project is a large search for gravitational lenses with the
VLA, with an explicit
goal of finding lens systems which can be used to measure $H_0$.  The
gravitational lens B1608+656 (RA: 16 09 13.956, Dec: +65 32 28.971,
J2000) was observed in the first phase of CLASS and was immediately
recognized as a lens system.  The radio discovery image shows four
unresolved components in a typical lens geometry (\cite{6_m1608}; see
Fig.~\ref{fig_1608map} for a map of the system).  The system was also
discovered in a search for gigahertz-peaked spectrum sources and was
found to be the lensed core of a classical radio double source
(\cite{6_s1608}).  Further investigations of the system have provided
data crucial for using the system for measuring $H_0$.  We have
measured the lens redshift ($z_{\ell} = 0.630$; \cite{6_m1608}) and
source redshift ($z_s = 1.394$; \cite{6_f1608}).  Optical and infrared
observations taken with the {\em Hubble Space Telescope} reveal that
the background source is being lensed by a pair of possibly merging
galaxies (\cite{6_hstlens}).  The positions of the
lensing galaxies relative to the lensed images are important
constraints on models of the lensing potential (\cite{6_lkmodel};
hereafter Paper 2).  The HST images also show arcs due to the lensing
of stellar emission from the background source.  These arcs could
be used as further constraints of the lens model.

Flat-spectrum cores of radio galaxies such as the lensed object in
B1608+656 are often variable.  To test for variability in the
B1608+656 background source, we made several observations of the
system with the VLA, separated by time-scales of months.  These data
showed that the flux density of the background source varied by up to
15\%.  The variability makes B1608+656 an excellent candidate for a
dedicated monitoring program to determine time delays.  This paper
presents the results of VLA monitoring from October 1996 to May 1997.
These observations have resulted in the measurement of the three
independent time delays in the B1608+656 system and a subsequent
determination of $H_0$.  The Hubble constant is expressed as $H_0 =
100\,h~{\rm km}\,{\rm s}^{-1}\,{\rm Mpc}^{-1}$.  Throughout this paper
we assume $(\Omega_M, \Omega_\Lambda) = (1,0)$.  The effect of varying
the cosmological model is treated in Paper 2.

\section{Observations}

We observed B1608+656 between 1996 October 10 and 1997 May 09, during
which time the VLA was in the A, BnA, and B configurations.  The 64
epochs were separated, on average, by 3.6~d.  The observations were
carried out at 8.5~GHz, giving angular resolutions ranging from
0\farcs25 to 0\farcs7 in the different array configurations.  The
observations are summarized in Table~\ref{tab_obs}.  The typical
observation is 60~min long and includes scans on B1608+656, a flux
calibrator (3C\,286 or 3C\,48), a phase calibrator (1642+689) chosen
from the VLA calibrator list (\cite{6_vlacalib}), and two secondary
flux calibrators (1634+627 and 1633+741).  The secondary flux
calibrators are nearby steep-spectrum sources that are not expected to
vary over the time-scales of the observations.  We observe these
sources to determine corrections for errors in the absolute flux
calibration from epoch to epoch.  The basic observing pattern is:

\begin{itemize} 
\item 1642+689 (1 min on source) 
\item 1634+627 (1 min on source) 
\item B1608+656 (4--6 min on source) 
\item 1633+741 (2 min on source) 
\end{itemize} 

A typical 60~min observation begins with a 9 -- 10~min scan (including
slew time) on the flux calibrator, contains three repetitions of the basic
pattern on B1608+656, and ends with scans on 1642+689 (1~min) and the
flux calibrator again ($\sim$3~min).  For the few observations that are
30~min or 90~min in length, the number of repetitions of the basic
pattern is altered.  

\begin{deluxetable}{rlclrrl}
\tablewidth{0pt}
\scriptsize
\tablecaption{Observations\label{tab_obs}}
\tablehead{
\colhead{}
 & \colhead{}
 & \colhead{Array}
 & \colhead{Start Time}
 & \colhead{$t_{tot}$}
 & \colhead{$t_{1608}$}
 & \\
\colhead{Epoch}
 & \colhead{MJD$-$50000}
 & \colhead{Configuration}
 & \colhead{(LST)}
 & \colhead{(min)}
 & \colhead{(min)}
 & \colhead{Comments}
}
\startdata
1996 Oct 10 & 366 & D$\rightarrow$A   & 16:30 & 60 & 27 & \\
1996 Oct 12 & 368 & D$\rightarrow$A   & 15:30 & 30 &  5 & \\
1996 Oct 16 & 372 & D$\rightarrow$A   & 21:00 & 60 & 21 & \\
1996 Oct 18 & 374 & A                 & 17:30 & 60 & 21 & \\
1996 Oct 20 & 376 & A                 & 01:30 & 60 & 21 & Elev. $\leq 30^{\circ}$.  Thunderstorms \\
1996 Oct 23 & 379 & A                 & 17:30 & 30 &  6 & \\
1996 Oct 26 & 382 & A                 & 14:30 & 60 & 18 & $T_{sys} > 100$K . \\
1996 Oct 29 & 385 & A                 & 22:30 & 30 &  5 & Elev. $\leq 30^{\circ}$. Gusting winds \\
1996 Oct 31 & 387 & A                 & 20:00 & 60 & 18 & \\
1996 Nov 01 & 388 & A                 & 18:30 & 60 & 18 & \\
1996 Nov 03 & 390 & A                 & 18:00 & 60 & 21 & Wind $\geq$ 10~m/s \\
1996 Nov 07 & 394 & A                 & 07:00 & 30 &  6 & Elev. $\leq 30^{\circ}$ \\
1996 Nov 08 & 395 & A                 & 13:00 & 60 & 21 & \\
1996 Nov 11 & 398 & A                 & 13:00 & 60 & 18 & \\
1996 Nov 12 & 399 & A                 & 07:30 & 30 &  8 & Elev. $\leq 30^{\circ}$ \\
1996 Nov 15 & 402 & A                 & 17:00 & 60 & 21 & High wind gusts \\ 
1996 Nov 18 & 405 & A                 & 15:30 & 60 & 21 & \\
1996 Nov 23 & 410 & A                 & 17:00 & 60 & 20 & Wind $\geq$ 10~m/s \\
1996 Nov 27 & 414 & A                 & 21:00 & 60 & 20 & \\
1996 Dec 01 & 418 & A                 & 14:00 & 60 & 20 & \\
1996 Dec 05 & 422 & A                 & 14:30 & 30 &  7 & \\
1996 Dec 07 & 424 & A                 & 15:00 & 30 &  5 & \\
1996 Dec 10 & 427 & A                 & 17:00 & 60 & 20 & Wind $\geq$ 10~m/s \\
1996 Dec 15 & 432 & A                 & 17:00 & 90 & 34 & \\
1996 Dec 20 & 437 & A                 & 19:15 & 30 &  5 & \\
1996 Dec 23 & 440 & A                 & 17:30 & 90 & 34 & \\
1996 Dec 24 & 441 & A                 & 15:30 & 30 & 10 & \\
1996 Dec 29 & 446 & A                 & 15:00 & 60 & 20 & \\
1997 Jan 03 & 451 & A                 & 16:30 & 60 & 20 & Wind $\geq$ 10~m/s \\
1997 Jan 07 & 455 & A                 & 18:30 & 90 & 35 & \\
1997 Jan 11 & 459 & A                 & 16:00 & 60 & 20 & \\
1997 Jan 16 & 464 & A$\rightarrow$BnA & 16:00 & 60 & 20 & \\
1997 Jan 17 & 465 & A$\rightarrow$BnA & 09:30 & 30 &  8 & Elev. $\leq 30^{\circ}$. Flurries \\
1997 Jan 20 & 468 & A$\rightarrow$BnA & 15:30 & 60 & 20 & \\
1997 Jan 26 & 474 & BnA               & 14:00 & 60 & 20 & Rain \\
1997 Jan 30 & 478 & BnA               & 13:00 & 60 & 19 & \\
1997 Feb 02 & 481 & BnA               & 20:00 & 60 & 20 & \\
1997 Feb 08 & 487 & BnA               & 19:00 & 60 & 19 & \\
1997 Feb 13 & 492 & BnA$\rightarrow$B & 15:30 & 60 & 18 & \\
1997 Feb 18 & 497 & B                 & 18:00 & 60 & 19 & \\
1997 Feb 23 & 502 & B                 & 18:00 & 60 & 19 & \\
1997 Feb 28 & 507 & B                 & 18:00 & 60 & 19 & Snow storms. \\
1997 Mar 03 & 510 & B                 & 18:00 & 60 & 19 & \\
1997 Mar 08 & 515 & B                 & 16:00 & 60 & 18 & \\
1997 Mar 14 & 521 & B                 & 18:15 & 75 & 30 & \\
1997 Mar 16 & 523 & B                 & 17:45 & 45 & 12 & \\
1997 Mar 21 & 528 & B                 & 18:00 & 60 & 20 & \\
1997 Mar 25 & 532 & B                 & 21:30 & 60 & 19 & Wind $\geq$ 10~m/s. Snowing. \\
1997 Mar 30 & 537 & B                 & 19:00 & 90 & 30 & \\
1997 Apr 01 & 539 & B                 & 17:30 & 60 & 20 & \\
1997 Apr 05 & 543 & B                 & 18:00 & 60 & 20 & \\
1997 Apr 07 & 545 & B                 & 18:00 & 60 & 20 & \\
1997 Apr 11 & 549 & B                 & 18:00 & 60 & 20 & \\
1997 Apr 15 & 553 & B                 & 18:00 & 60 & 20 & \\
1997 Apr 19 & 557 & B                 & 15:00 & 60 & 18 & \\
1997 Apr 22 & 560 & B                 & 18:00 & 60 & 20 & \\
1997 Apr 26 & 564 & B                 & 18:00 & 60 & 20 & \\
1997 May 03 & 571 & B                 & 18:00 & 60 & 20 & \\
1997 May 08 & 576 & B                 & 18:00 & 90 & 30 & \\
1997 May 13 & 581 & B                 & 18:00 & 60 & 20 & \\
1997 May 17 & 585 & B                 & 16:30 & 60 & 19 & \\
1997 May 21 & 589 & B                 & 18:00 & 60 & 20 & \\
1997 May 23 & 591 & B                 & 18:00 & 60 & 20 & \\
1997 May 26 & 594 & B                 & 20:00 & 60 & 19 & \\ 
\enddata
\end{deluxetable}

\section{Data Reduction}

\subsection{Calibration}

The data reduction is separated into two major steps, calibration and
mapping.  The data for each epoch are calibrated using standard
routines in the NRAO data reduction package AIPS.  Before calibration,
the data quality for all sources is assessed and bad points are
flagged with the EDITA and TVFLG tasks.  Both of the flux calibrators
are heavily resolved by the VLA in A configuration at 8.5~GHz.  Hence,
we cannot treat the calibrators as point sources without limiting the
number of baselines that can be used to calculate phase and gain
solutions.  In order to increase the number of baselines available for
the calculations, we create models of 3C\,286 and 3C\,48 that
incorporate the extended emission from the sources.  We combine
observations from several epochs with the DBCON task in AIPS.  The
resulting data sets have excellent $(u,v)$-plane coverage, from which
we can make high dynamic-range maps.  The mapping, which is performed
in the DIFMAP package (\cite{6_difmap}), consists of alternating
iterations of CLEANing (\cite{6_clean}) and self-calibration.  The
final lists of CLEAN components are read back into AIPS and serve as
the calibrator models.

The procedures for the phase calibrator are simpler because the
emission from 1642+689 is dominated by an unresolved component.  Thus,
the assumption that 1642+689 is a point source leads to adequate phase
and gain solutions.  These calibration solutions are applied to the
B1608+656, 1633+741, and 1634+627 data.

\subsection{Source Maps\label{smaps}}

We map the data and determine flux densities using the DIFMAP package.
We do not expect to see any structural changes over the course of the
observations; only changes in flux densities should be observed.  To
treat the data from each epoch in a uniform fashion and to shorten the
mapping procedure, we create models of the observed source structures
from high dynamic-range maps.  For epochs with noisy data, these
models are needed to fix the locations of the regions of low surface
brightness emission, which otherwise would not be well-constrained by
the data.

\begin{figure}
\plotone{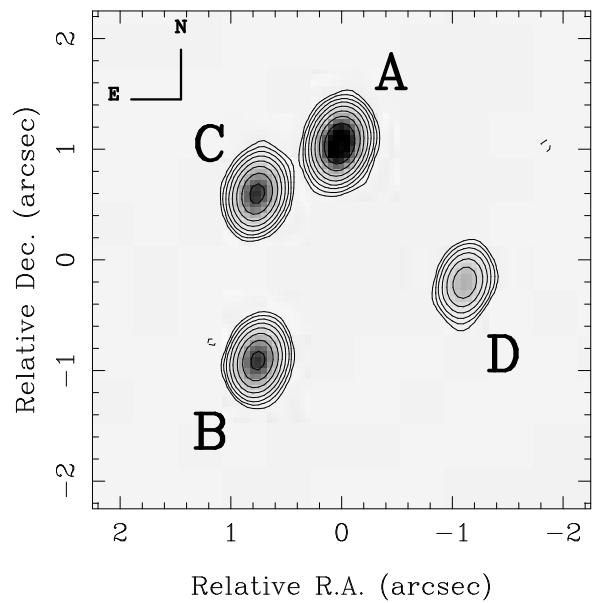}
\caption
{Map of B1608+656 from observation on
1996 November 18.  The contours are ($-$3, 3, 6, 12, 24, 48, 96,
192, 384, 768) times the RMS noise level of 0.035~mJy/beam.  The map is
made by fitting point source components to the $(u,v)$ data and
restoring with a 0\farcs33$\times$0\farcs23 restoring beam.
\label{fig_1608map}}
\end{figure}

\begin{figure}
\plotone{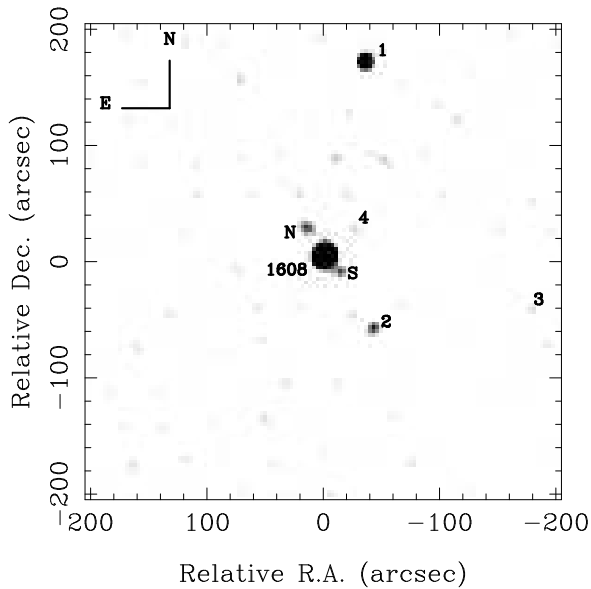}
\caption
{The field of B1608+656, showing the lens system and nearby radio sources.
The object labelled ``1608'' consists of the four components shown in 
Fig.~\ref{fig_1608map}.  Objects ``N'' and ``S'' correspond to the 
northern and southern radio lobes of B1608+656 seen in low frequency maps 
(\cite{6_s1608}).
\label{fig_1608field}}
\end{figure}

To make the high dynamic-range maps of each source, we combine 13
high-quality data sets from A and B configuration observations with
the DBCON task.  The combined data sets, which have excellent $(u,v)$
coverage, are then mapped in DIFMAP.  All the maps are made with
natural weighting.  The secondary flux calibrators, which have
significant emission from extended structures, are mapped by using an
iterative cycle of CLEANing and self-calibration.  Both phase and
amplitude self-calibration are used.  The models for these sources
consist of the final lists of CLEAN components.  The emission from
B1608+656 is dominated by the four unresolved images of the background
source (Fig.~\ref{fig_1608map}).  Hence, instead of CLEANing the data,
we assign point-source model components to the four images of the
background source.  We then use the DIFMAP {\tt modelfit} function,
which varies the component positions and flux densities to obtain the
best fit to the $(u,v)$-plane visibilities.  The model fitting
iterations are alternated with phase and amplitude self-calibration.
In later rounds of the model fitting, several nearby weak sources are
seen in the residual maps.  These sources are included in the model
for the last few iterations of the model fitting.  The nearby sources
can be seen in Fig.~\ref{fig_1608field}; their locations and flux
densities are listed in Table~\ref{tab_posflux}.

\begin{deluxetable}{lrrr}
\tablewidth{0pt}
\scriptsize
\tablecaption{Object Positions and Flux Densities 
\label{tab_posflux}}
\tablehead{
\colhead{Object}
 & \colhead{$\alpha$ (\arcsec)\tablenotemark{a}}
 & \colhead{$\delta$ (\arcsec)\tablenotemark{a}}
 & \colhead{$S_{8.4}$ (mJy)\tablenotemark{b}}
}
\startdata
B1608+656 A &   $+$0.018           &   $+$1.001           & 34.3 \\
B1608+656 B &   $+$0.757           &   $-$0.961           & 16.8 \\
B1608+656 C &   $+$0.763           &   $+$0.548           & 17.4 \\
B1608+656 D &   $-$1.111           &   $-$0.256           &  5.9 \\
B1608+656 N &  $+$15.0\phantom{00} &  $+$23.6\phantom{00} &  0.6 \\
B1608+656 S &  $-$13.9\phantom{00} &  $-$13.0\phantom{00} &  0.5 \\
Source 1    &  $-$34.7\phantom{00} & $+$167.0\phantom{00} &  3.4 \\
Source 2    &  $-$42.0\phantom{00} &  $-$61.2\phantom{00} &  0.8 \\
Source 3    & $-$178.0\phantom{00} &  $-$43.5\phantom{00} &  0.2 \\
Source 4    &  $-$26.7\phantom{00} &  $+$24.6\phantom{00} &  0.2 \\
\enddata
\tablenotetext{a}{Positions relative to map center at RA: 16:09:13.9530,  
Dec: +65:32:27.998 (J2000).}
\tablenotetext{b}{Flux densities for B1608+656 components A -- D are the
mean flux densities over the period of VLA monitoring (see \S\ref{flat}).}
\end{deluxetable}

The first step in the mapping procedure at each epoch is to read in
the data and perform a phase-only self-calibration against the model
of the source.  This procedure aligns the phase center of the
observation with that of the model and eliminates the need for many
early steps of cleaning and self-calibration.  After this point, the
procedures used for B1608+656 differ from those used for the secondary
flux calibrators, as discussed below.

\subsubsection{B1608+656\label{sec_1608maps}}

For each epoch, we determine the flux densities of the four lensed
images in the B1608+656 system as follows.  After the initial phase
self-calibration step, the model described in the previous section is
varied to find the best fit to the $(u,v)$-plane visibilities for that
epoch.  The component positions are held fixed and only the flux
densities are allowed to vary.  After several iterations of model
fitting, another phase self-calibration is performed against the new
model and more iterations of the model fitting are performed.  At this
point, the component flux densities and the RMS noise in the residual
map are recorded in a log file (the ``phasecal'' data set).  We then
perform an overall gain calibration on the data, getting one gain
correction per antenna for the observation.  Typical gain corrections
are on the order of 1 -- 2\%.  Another round of model fitting is
performed and the final component flux densities and RMS noise in the
residual map are recorded in the log file (the ``gscale'' data set).
We record the two separate data sets in case the gain calibration
introduces any errors which may bias the subsequent analysis.  All
subsequent analysis is performed on both data sets and no significant
differences are seen in the results.

\subsubsection{Secondary Flux Calibrators}

The steep-spectrum secondary flux-density calibrators contain
significant extended emission and should not vary over the course of
the observations.  We expect that any observed variations in flux
density are due to errors in the absolute flux calibration, and as
such can be expressed as an overall scaling of the model CLEAN
component flux densities.  That is, the CLEAN component flux densities
should not vary with respect to each other.  The task of finding the
overall flux density of these sources is thus simplified into finding
the scale factor $(\mu)$ that, when multiplied by the component flux
densities, gives the best fit to the data.  To find the best-fit
scaling for each data set, we create 11 scaled model files, based on
the CLEAN-component models described above.  The 11 files
have $\mu$ ranging from 0.9 to 1.1 in steps of 0.02.  Each of the
scaled models is compared to the $(u,v)$-plane visibilities and a
reduced $\chi^2$ goodness-of-fit value is returned.  We then fit a
parabola to the points in the reduced-$\chi^2$ curve and find the
value of $\mu$ which corresponds to the minimum reduced $\chi^2$.  This
scaling gives the total flux density of the source at that epoch.

\subsection{Light Curve Editing}

\begin{figure}
\plotone{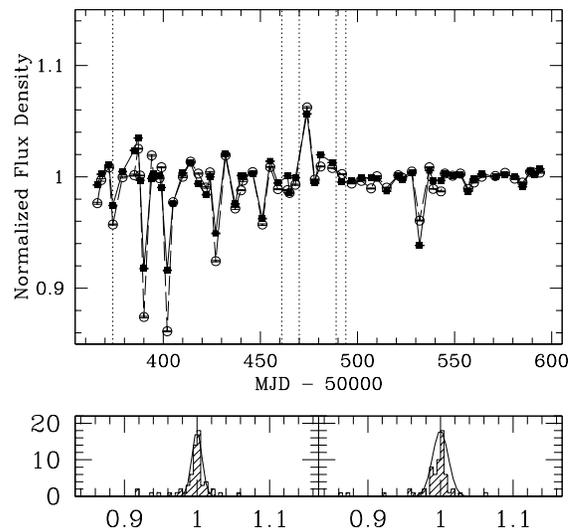}
\caption
{\label{fig_normcmp}{\bf Top}-- Light curves of the secondary calibrators
1633+741 (open circles) and 1634+627 (filled boxes). Each curve has
been divided by its median value over the length of the observations.
The dotted vertical lines represent changes of array configuration.
See Table~\ref{tab_obs} for the array configuration at each epoch.
{\bf Bottom} -- The distribution of normalized flux densities about the
median values for 1634+627 (left) and 1633+741 (right).}
\end{figure}

Moore \& Hewitt (1997), in their analysis of the 15~GHz light curves
of the gravitational lens MG~0414+0534, developed objective criteria to
flag questionable data.  They deleted from their light curves all
points associated with observations with the following conditions: the
telescope elevations were less than 30$^{\circ}$, the wind speed was
greater than 10~m/s, or there was precipitation.  We have noted all
epochs satisfying their criteria in the ``Comments'' column of
Table~\ref{tab_obs}.  However, we are able to include many of these
points in our analysis because observations at 8.5~GHz are less
sensitive to the observing conditions than are observations taken at
15~GHz.  As such, we have only excluded epochs for which the data are
severely affected by the observing conditions.  This assessment is
made by examining the light curves of the secondary calibrator
sources.  All epochs for which the flux densities of the calibrators
deviate by more than 15\% from the mean value are deleted.  Only two
days are deleted after the application of this criterion: 376 and
382 (MJD$-$50000).  Note that epoch 376 satisfies two of the flagging
criteria defined by Moore \& Hewitt.  At epoch 382, the system
temperatures for all of the telescopes were in the range 100 -- 200 K,
as compared to the 30 -- 50 K system temperatures measured for all
other epochs.  These high system temperatures may have resulted from
the fact the subreflectors of several of the antennas had frozen prior
to the observation and had just thawed.  The signal-to-noise ratio of
the maps for epoch 382 were so low that no useful information could be
extracted from them.  The final edited light curves contain 62 epochs,
with an average spacing of 3.7~d.  Edited versions of the secondary
flux calibrator and B1608+656 light curves are shown in
Figures~\ref{fig_normcmp} and
\ref{fig_w08}.

\subsection{Secondary Flux Calibration and Final Light Curves\label{flat}}

The normalized light curves of the secondary flux calibrators are
shown in Fig.~\ref{fig_normcmp}.  As expected for non-variable
sources, the light curves are close to constant.  The scatter about
the mean is small, with the exception of a few outlying points.  All
of these outlier points are due to bad observing conditions.  The five
worst points, for example, occur on days 390, 402, 427, 474, and 532,
all of which experienced high winds, precipitation, or both
(Table~\ref{tab_obs}).  Because of these outliers, the formal
calculation of $\sigma$ overestimates the width of the error
distribution.  Consequently, we have estimated $\sigma$ using the
interquartile range (IQR), which is less affected by the outliers.  For a
Gaussian distribution, $\sigma = 0.741 * {\rm IQR}$.  For the
secondary flux calibrators, we obtain $\sigma = 0.7\%$ and $\sigma = 1.1\%$
for 1634+627 and 1633+741, respectively.  Gaussian distributions with these
values of $\sigma$ are reasonable fits to the data (see Fig.~\ref{fig_normcmp}).

\begin{figure}
\plotone{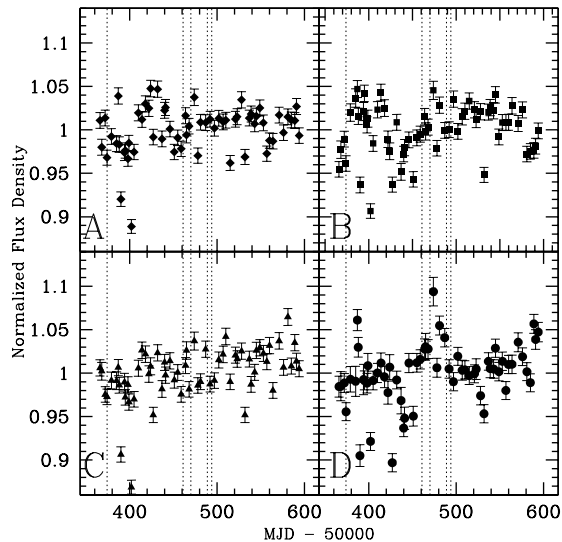}
\caption
{\label{fig_w08} Light curves for the four images of the background
source in B1608+656 {\em before} the secondary flux calibration has been
applied.  The light curves have been normalized by their mean values such
that equal fractional variations in the flux densities of the components will
have the same heights in the curves.}
\end{figure}

\begin{figure}
\plotone{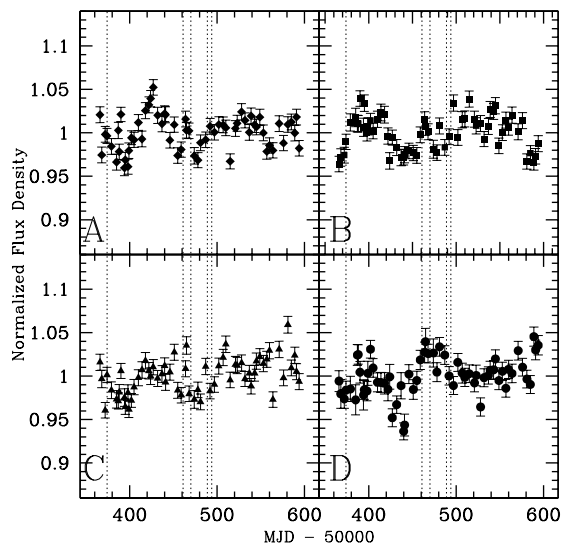}
\caption
{\label{fig_wflat08} Same as Fig.~\ref{fig_w08} but {\em after}
the application of the secondary flux calibration.}
\end{figure}

The light curves of the two secondary flux calibrators track
each other extremely well, suggesting that the observed variations are
due to errors in the absolute flux calibration of the data rather than
any intrinsic variability of the sources.  Because of this, we can use
the light curves of the secondary flux calibrators to remove residual
flux calibration errors in the B1608+656 component light curves.  We
create a calibration correction curve by first normalizing the light
curves of the secondary flux calibrators and then averaging the two
normalized fluxes at each epoch.  We then apply the secondary flux
calibration by dividing the B1608+656 component light curves by the
calibration correction curve.  The correction is effective in reducing
the scatter in the B1608+656 light curves, as can be seen by comparing
Fig.~\ref{fig_w08} to Fig.~\ref{fig_wflat08}.  The corrected component
light curves are listed in Table~\ref{tab_complcurve}.  An electronic
version of the light curve table may also be obtained from {\tt
http://www.nrao.edu/\~\,cfassnac/1608flux.tab}.

\begin{deluxetable}{ccccc}
\tablewidth{0pt}
\scriptsize
\tablecaption{Final Component Light Curves
\label{tab_complcurve}}
\tablehead{
\colhead{}
 & \multicolumn{4}{c}{Flux Density (mJy)}
 \\
\colhead{MJD$-$50000}
 & \colhead{A}
 & \colhead{B}
 & \colhead{C}
 & \colhead{D}
}
\startdata
366 & 34.831 $\pm$ 0.309 & 16.106 $\pm$ 0.149 & 17.602 $\pm$ 0.162 &  5.822 $\pm$ 0.070  \\
368 & 33.251 $\pm$ 0.307 & 16.233 $\pm$ 0.166 & 17.269 $\pm$ 0.175 &  5.736 $\pm$ 0.097  \\
372 & 34.063 $\pm$ 0.308 & 16.277 $\pm$ 0.151 & 16.643 $\pm$ 0.154 &  5.701 $\pm$ 0.063  \\
374 & 34.003 $\pm$ 0.294 & 16.532 $\pm$ 0.146 & 17.349 $\pm$ 0.153 &  5.760 $\pm$ 0.060  \\
379 & 33.589 $\pm$ 0.307 & 16.904 $\pm$ 0.165 & 17.042 $\pm$ 0.166 &  5.768 $\pm$ 0.085  \\
385 & 32.972 $\pm$ 0.310 & 17.002 $\pm$ 0.176 & 16.863 $\pm$ 0.175 &  5.695 $\pm$ 0.101  \\
387 & 34.217 $\pm$ 0.317 & 16.883 $\pm$ 0.161 & 16.831 $\pm$ 0.160 &  5.997 $\pm$ 0.070  \\
388 & 33.381 $\pm$ 0.299 & 16.884 $\pm$ 0.155 & 17.002 $\pm$ 0.156 &  6.000 $\pm$ 0.066  \\
390 & 34.845 $\pm$ 0.283 & 17.365 $\pm$ 0.149 & 17.429 $\pm$ 0.150 &  5.881 $\pm$ 0.073  \\
394 & 32.744 $\pm$ 0.307 & 16.811 $\pm$ 0.174 & 16.898 $\pm$ 0.175 &  5.727 $\pm$ 0.101  \\
395 & 33.060 $\pm$ 0.297 & 17.276 $\pm$ 0.159 & 16.715 $\pm$ 0.154 &  5.763 $\pm$ 0.064  \\
398 & 32.795 $\pm$ 0.295 & 16.705 $\pm$ 0.154 & 16.995 $\pm$ 0.157 &  5.755 $\pm$ 0.065  \\
399 & 33.422 $\pm$ 0.304 & 16.830 $\pm$ 0.162 & 16.661 $\pm$ 0.161 &  5.874 $\pm$ 0.082  \\
402 & 33.927 $\pm$ 0.271 & 16.943 $\pm$ 0.139 & 16.840 $\pm$ 0.138 &  6.036 $\pm$ 0.060  \\
405 & 33.846 $\pm$ 0.296 & 16.740 $\pm$ 0.150 & 17.107 $\pm$ 0.153 &  5.910 $\pm$ 0.062  \\
410 & 34.527 $\pm$ 0.310 & 16.966 $\pm$ 0.156 & 17.294 $\pm$ 0.159 &  5.813 $\pm$ 0.064  \\
414 & 33.878 $\pm$ 0.308 & 17.096 $\pm$ 0.160 & 17.447 $\pm$ 0.163 &  5.813 $\pm$ 0.067  \\
418 & 34.996 $\pm$ 0.315 & 17.054 $\pm$ 0.159 & 17.632 $\pm$ 0.164 &  5.807 $\pm$ 0.071  \\
422 & 35.218 $\pm$ 0.315 & 16.645 $\pm$ 0.157 & 17.431 $\pm$ 0.164 &  5.765 $\pm$ 0.077  \\
424 & 35.464 $\pm$ 0.323 & 16.168 $\pm$ 0.158 & 17.324 $\pm$ 0.168 &  5.850 $\pm$ 0.084  \\
427 & 35.909 $\pm$ 0.302 & 16.613 $\pm$ 0.144 & 17.505 $\pm$ 0.151 &  5.574 $\pm$ 0.061  \\
432 & 34.811 $\pm$ 0.317 & 16.423 $\pm$ 0.152 & 17.280 $\pm$ 0.160 &  5.663 $\pm$ 0.059  \\
437 & 34.475 $\pm$ 0.308 & 16.234 $\pm$ 0.158 & 17.368 $\pm$ 0.167 &  5.791 $\pm$ 0.088  \\
440 & 34.875 $\pm$ 0.310 & 16.236 $\pm$ 0.147 & 17.532 $\pm$ 0.158 &  5.483 $\pm$ 0.057  \\
441 & 34.831 $\pm$ 0.315 & 16.286 $\pm$ 0.155 & 17.203 $\pm$ 0.163 &  5.525 $\pm$ 0.074  \\
446 & 33.832 $\pm$ 0.304 & 16.358 $\pm$ 0.151 & 17.405 $\pm$ 0.160 &  5.868 $\pm$ 0.064  \\
451 & 34.444 $\pm$ 0.298 & 16.320 $\pm$ 0.146 & 17.802 $\pm$ 0.158 &  5.765 $\pm$ 0.066  \\
455 & 33.230 $\pm$ 0.301 & 16.271 $\pm$ 0.150 & 17.042 $\pm$ 0.157 &  5.825 $\pm$ 0.062  \\
459 & 33.463 $\pm$ 0.298 & 16.667 $\pm$ 0.152 & 16.946 $\pm$ 0.154 &  5.964 $\pm$ 0.064  \\
464 & 34.451 $\pm$ 0.310 & 16.851 $\pm$ 0.156 & 17.364 $\pm$ 0.160 &  5.979 $\pm$ 0.068  \\
465 & 34.010 $\pm$ 0.309 & 16.720 $\pm$ 0.164 & 17.825 $\pm$ 0.173 &  6.049 $\pm$ 0.090  \\
468 & 33.989 $\pm$ 0.305 & 16.624 $\pm$ 0.152 & 16.867 $\pm$ 0.154 &  5.970 $\pm$ 0.063  \\
474 & 33.015 $\pm$ 0.322 & 16.290 $\pm$ 0.172 & 16.755 $\pm$ 0.176 &  5.974 $\pm$ 0.095  \\
478 & 32.845 $\pm$ 0.295 & 16.223 $\pm$ 0.149 & 16.945 $\pm$ 0.156 &  5.845 $\pm$ 0.064  \\
481 & 33.523 $\pm$ 0.306 & 16.735 $\pm$ 0.155 & 16.709 $\pm$ 0.155 &  6.015 $\pm$ 0.063  \\
487 & 33.643 $\pm$ 0.306 & 16.325 $\pm$ 0.151 & 17.410 $\pm$ 0.161 &  5.960 $\pm$ 0.062  \\
492 & 34.166 $\pm$ 0.308 & 16.528 $\pm$ 0.152 & 16.923 $\pm$ 0.155 &  5.820 $\pm$ 0.062  \\
497 & 33.754 $\pm$ 0.304 & 17.080 $\pm$ 0.156 & 16.961 $\pm$ 0.155 &  5.725 $\pm$ 0.060  \\
502 & 34.051 $\pm$ 0.307 & 16.427 $\pm$ 0.150 & 17.323 $\pm$ 0.158 &  5.879 $\pm$ 0.060  \\
507 & 34.040 $\pm$ 0.308 & 16.761 $\pm$ 0.156 & 17.495 $\pm$ 0.162 &  5.808 $\pm$ 0.067  \\
510 & 33.924 $\pm$ 0.307 & 16.779 $\pm$ 0.154 & 17.754 $\pm$ 0.163 &  5.779 $\pm$ 0.061  \\
515 & 32.624 $\pm$ 0.292 & 17.156 $\pm$ 0.156 & 17.050 $\pm$ 0.155 &  5.804 $\pm$ 0.060  \\
521 & 33.901 $\pm$ 0.307 & 16.774 $\pm$ 0.154 & 17.356 $\pm$ 0.159 &  5.743 $\pm$ 0.060  \\
523 & 34.102 $\pm$ 0.310 & 16.648 $\pm$ 0.157 & 17.337 $\pm$ 0.162 &  5.792 $\pm$ 0.069  \\
528 & 34.548 $\pm$ 0.314 & 16.691 $\pm$ 0.155 & 17.386 $\pm$ 0.161 &  5.582 $\pm$ 0.061  \\
532 & 34.210 $\pm$ 0.295 & 16.394 $\pm$ 0.145 & 17.069 $\pm$ 0.150 &  5.777 $\pm$ 0.061  \\
537 & 33.749 $\pm$ 0.307 & 16.636 $\pm$ 0.153 & 17.181 $\pm$ 0.158 &  5.792 $\pm$ 0.058  \\
539 & 34.388 $\pm$ 0.309 & 16.958 $\pm$ 0.155 & 16.927 $\pm$ 0.155 &  5.826 $\pm$ 0.061  \\
543 & 34.049 $\pm$ 0.306 & 16.921 $\pm$ 0.155 & 17.185 $\pm$ 0.157 &  5.830 $\pm$ 0.062  \\
545 & 33.959 $\pm$ 0.308 & 17.031 $\pm$ 0.156 & 17.422 $\pm$ 0.160 &  5.903 $\pm$ 0.060  \\
549 & 34.338 $\pm$ 0.311 & 16.267 $\pm$ 0.150 & 17.507 $\pm$ 0.161 &  5.758 $\pm$ 0.061  \\
553 & 33.728 $\pm$ 0.306 & 16.519 $\pm$ 0.152 & 17.374 $\pm$ 0.160 &  5.820 $\pm$ 0.061  \\
557 & 33.004 $\pm$ 0.295 & 16.755 $\pm$ 0.152 & 17.459 $\pm$ 0.158 &  5.707 $\pm$ 0.060  \\
560 & 33.241 $\pm$ 0.300 & 16.622 $\pm$ 0.152 & 17.630 $\pm$ 0.161 &  5.833 $\pm$ 0.061  \\
564 & 33.049 $\pm$ 0.300 & 16.850 $\pm$ 0.156 & 16.666 $\pm$ 0.154 &  5.805 $\pm$ 0.062  \\
571 & 34.091 $\pm$ 0.309 & 16.535 $\pm$ 0.152 & 17.653 $\pm$ 0.162 &  5.957 $\pm$ 0.063  \\
576 & 33.332 $\pm$ 0.302 & 16.748 $\pm$ 0.154 & 17.093 $\pm$ 0.157 &  5.847 $\pm$ 0.059  \\
581 & 34.055 $\pm$ 0.308 & 15.971 $\pm$ 0.147 & 18.138 $\pm$ 0.166 &  5.768 $\pm$ 0.060  \\
585 & 34.151 $\pm$ 0.307 & 16.127 $\pm$ 0.148 & 17.287 $\pm$ 0.158 &  5.732 $\pm$ 0.060  \\
589 & 33.723 $\pm$ 0.307 & 15.952 $\pm$ 0.148 & 17.538 $\pm$ 0.162 &  6.051 $\pm$ 0.064  \\
591 & 34.341 $\pm$ 0.312 & 16.068 $\pm$ 0.150 & 17.214 $\pm$ 0.160 &  5.962 $\pm$ 0.066  \\
594 & 33.130 $\pm$ 0.302 & 16.309 $\pm$ 0.152 & 17.018 $\pm$ 0.158 &  5.993 $\pm$ 0.065  \\
\enddata
\end{deluxetable}

The errors on the final flux densities in Table~\ref{tab_complcurve}
are a combination of additive and multiplicative terms.  The additive
uncertainty is well-approximated by the RMS noise in the residual map
at the end of the model-fitting.  The multiplicative term is
indicative of how well the model-fitting procedure is able to find the
``true'' total flux density of a source.  We estimate this error by
calculating the ratio of the flux densities of 1634+627 and 1633+741,
which we expect not to vary with respect to each other.  The flux
density ratio is not affected by errors in the absolute flux
calibration, so any scatter in the ratio can be attributed to errors
in the model fitting procedure.  From the scatter in the flux density
ratio, we estimate the fractional error in the component flux
densities to be 0.9\%; the product of the component flux densities and
this value gives the multiplicative uncertainty.  The final errors on
the flux densities are a combination in quadrature of the additive and
multiplicative terms.  For the three brightest components (A, B, and
C), the multiplicative terms dominate; for component D the
multiplicative and additive terms are comparable.

The flux density of a lensed image of the background source is $\mu_i
\,S_{\nu}$, where $S_{\nu}$ is the true flux density of the background source and
$\mu_i$ is the magnification associated with the $i^{th}$ (unresolved)
image.  Similarly, the lensed images will vary by $\mu_i\,dS_{\nu}$ if
the flux density of the background source changes by an amount
$dS_{\nu}$.  Hence, any variations in the B1608+656 component light
curves that are due to variations in the background source must have
the same {\em fractional} amplitudes in all four curves.  For this
reason, the component light curves presented in Figs.~\ref{fig_w08}
and \ref{fig_wflat08} are divided by their mean flux densities to
allow direct comparison of the fractional variations in the curves.
The mean values used to normalize the light curves are 34.29, 16.79,
17.41, and 5.884~mJy for components A, B, C, and D, respectively.  All
four light curves show variations in flux density at the $\sim$5\%
level, and there are common features that can be seen in each.  Each
light curve shows a rise of $\sim$5\% in flux density, followed by a
plateau lasting $\sim$20~d and then a drop of $\sim$4\%.  

\section{Determination of Time Delays\label{timedelay}}

Models of the B1608+656 system predict that, if the background source
is variable, the lensed images will vary in the order B $\rightarrow$
A $\rightarrow$ C $\rightarrow$ D (\cite{6_m1608}; Paper 2).  This is,
in fact, the behavior seen in the B1608+656 light curves.  In order to
determine time delays between component B and the other three
components, we have used three statistical methods: (1) smoothing and
$\chi^2$ minimization, (2) smoothing and cross-correlation, and (3)
dispersion analysis (Pelt et al.\ 1994, 1996, 1998).  The methods are
discussed below.  All of the analysis has been performed on both the
``phasecal'' and ``gscale'' data sets (\S\ref{sec_1608maps}).  No
significant differences in the results are seen, so we present only
the results from the ``gscale'' set in the subsequent discussion.

\subsection{Methods Using Smoothing/Interpolation\label{smooth}}

The first two methods of determining the time delays from the B1608+656
light curves require that the observed data be transferred onto a
regular grid.  Some previous determinations of time delays (e.g.,
\cite{6_k0957_2}; \cite{6_b0218}) have accomplished this transfer by
linear interpolation of the data.  However, the flux density
variations seen in the B1608+656 images are small compared to the noise
in the curves; therefore, linear interpolation can amplify noise
spikes.  In contrast, smoothing reduces the effects of noise compared
to the true variations, if the variations have typical time-scales
significantly longer than the sampling interval.  We smooth and
re-sample the data by calculating the weighted mean of points within a
smoothing window that is moved from the beginning to the end of the
observations in regular steps.  The step size is set to one day.

We smooth each light curve with several different functions to avoid
biasing the results by our choice of weighting function or window
size.  For completeness, we also include the results obtained from
interpolating the data by piecewise linear interpolation.  We use the
following smoothing schemes: (1) boxcar-weighted mean, (2)
triangle-weighted mean, (3) Gaussian-weighted mean, (4)
boxcar-weighted mean with a variable-width smoothing window, and (5)
triangle-weighted mean with a variable-width window.  In the last two
schemes, the width of the smoothing window is varied such that the
same number of points are always included in the window.  We use 3, 5,
and 7 point windows for these methods.  For the fixed-width window
schemes, we smooth with window widths of 5, 10, and 15~d.  The
Gaussian smoothing scheme uses values of $\sigma$ equal to 3, 5, and
7~d.  In addition to multiplying by the value of the smoothing
function, the data points are also variance-weighted.  Thus, the
overall weighting on a point at $t = t_i$ which is being used to
calculate a weighted mean at $t = t_k$ is:
\begin{equation}
w_i = \frac{\frac{1}{\sigma_i^2}\,g(t_i - t_k)}
{\sum_j [\frac{1}{\sigma_j^2}\,g(t_j-t_k)]},
\end{equation}
where $g(t)$ is the value of the smoothing function (e.g., g(t) = 1
for boxcar smoothing), $\sigma_i$ is the uncertainty on the flux
density at $t = t_i$ (calculated in \S\ref{flat}), and the sum is over
the points used to calculate the mean.  The variance of the weighted
mean $y$ at $t = t_k$ in the smoothed curve then becomes
\begin{equation}
\sigma_{y,k}^2 = 
\frac{1}{\left ( \sum_i [\frac{1}{\sigma_i^2}\,g(t_i - t_k)] \right )^2}
\sum_i \frac{[g(t_i - t_k)]^2}{\sigma_i^2},
\label{eqn_var_wmean}
\end{equation}
and $\sigma_{y,k}$ is taken as the uncertainty in the smoothed
flux density at that step.

\subsubsection{$\chi^2$ Minimization\label{chi2}}

For the $\chi^2$ minimization technique, we compare two light curves,
one of which is designated the ``control'' curve and the other of
which is designated the ``comparison'' curve.  Both curves are
smoothed, and then the comparison curve is multiplied by a scale
factor $\mu$ so that its mean flux density is comparable to that of
the control curve.  After this scaling, the comparison curve is
shifted in time by an amount $\tau$ with respect to the control curve.
We form a grid of $(\mu, \tau)$ pairs and calculate the $\chi^2$
statistic at each grid point.  The step size for $\tau$ used in the
grid is 1~d and the minimum and maximum shifts are set to $\pm$114~d,
i.e., half of the total length of the observations.  The amplitude
scale factors are set as percentages of the scale factor $\mu_0$ that
equalizes the mean flux densities of the two curves being compared.
The scale factors used in the grid range from 0.9$\mu_0$ to 1.1$\mu_0$,
in steps of 0.001$\mu_0$.  

One result of the smoothing and interpolation performed on the input
light curves is that the points in the interpolated curves are no
longer independent.  When computing the reduced $\chi^2$ statistic we
estimate an effective number of independent points, $N_{eff}$, in the
overlapping region between the shifted and unshifted curves.  This
quantity is set to the number of points that would have been present
in the overlapping region if the original light curves had been
regularly sampled at the average spacing of 3.7~d.  This simplifying
assumption should not significantly affect the quantity in which we
are interested, which is the location of the minimum on the $\chi^2$
surface.  In fact, the $\chi^2$-minimization method finds delays that
match closely those found from the completely non-interpolative
dispersion method (\S\ref{dispersion}), indicating the robustness of
the delays determined from this method.

\begin{figure}
\plotone{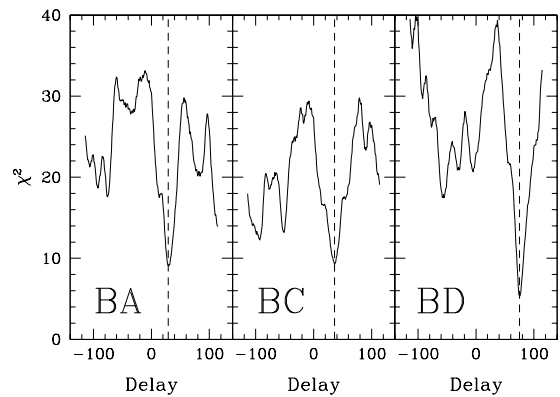}
\caption
{\label{fig_pchi3} Plots of reduced $\chi^2$ vs. lag from comparison
of the light curves of components B and A (left), B and C (middle),
and B and D (right). Each light curve is smoothed with a boxcar of
width 10~d before the $\chi^2$ curves are calculated.  The minima
(vertical dashed lines) are at lags of 29.3, 36.2, and 75.3~d,
respectively.}
\end{figure}

We repeat the $\chi^2$ minimization three times, once for each
independent pair of light curves.  The component B light curve is
always taken as the control curve.  In the three repetitions of the
$\chi^2$ minimization, the comparison curves are A, C, and D,
respectively.  The minimum delay is found by fitting a parabola to the
points at the minimum of the gridded $\chi^2$ curve.
Table~\ref{tab_delays} summarizes the results.  Typical
goodness-of-fit curves for the three pairs of light curves are shown
in Figure~\ref{fig_pchi3}.

\begin{deluxetable}{lrrrrrrr}
\tablewidth{0pt}
\scriptsize
\tablecaption{Time Delays from $\chi^2$ Minimization and
Cross-Correlation
\label{tab_delays}}
\tablehead{
\colhead{Smoothing} 
 & \colhead{Smoothing}
 & \multicolumn{3}{c}{$\chi^2$ Minimization}
 & \multicolumn{3}{c}{Cross-correlation}
 \\
\colhead{Function}
 & \colhead{``Width''\tablenotemark{a}}
 & \colhead{$\Delta t_{BA}$}
 & \colhead{$\Delta t_{BC}$}
 & \colhead{$\Delta t_{BD}$}
 & \colhead{$\Delta t_{BA}$}
 & \colhead{$\Delta t_{BC}$}
 & \colhead{$\Delta t_{BD}$}
}
\startdata
Boxcar         &  5      &  30.97 & 35.96 & 74.27 & 28 & 36 & 74  \\ 
               & 10      &  29.25 & 36.24 & 75.32 & 29 & 37 & 75  \\ 
               & 15      &  31.84 & 36.10 & 75.58 & 32 & 36 & 75  \\ 
Triangle       &  5      &  29.16 & 36.07 & 74.20 & 28 & 36 & 74  \\ 
               & 10      &  29.35 & 35.63 & 75.27 & 29 & 35 & 75  \\ 
               & 15      &  30.77 & 36.03 & 75.97 & 30 & 36 & 76  \\ 
Gaussian       &  3      &  30.65 & 35.93 & 75.93 & 30 & 36 & 76  \\ 
               &  5      &  31.55 & 36.17 & 76.70 & 31 & 37 & 76  \\ 
               &  7      &  31.52 & 36.37 & 77.06 & 31 & 37 & 77  \\ 
Variable-width &  3      &  30.13 & 35.79 & 76.58 & 30 & 36 & 75  \\ 
boxcar         &  5      &  29.92 & 36.18 & 77.32 & 30 & 37 & 78  \\ 
               &  7      &  29.44 & 36.38 & 77.22 & 30 & 39 & 77  \\ 
Variable-width &  3      &  30.52 & 35.84 & 76.48 & 30 & 36 & 76  \\ 
triangle       &  5      &  30.74 & 36.22 & 77.03 & 30 & 37 & 77  \\ 
               &  7      &  29.93 & 36.25 & 77.21 & 30 & 37 & 77  \\ 
Interpolation  & \nodata &  29.02 & 36.09 & 74.82 & 28 & 36 & 74  \\ 
\enddata
\tablenotetext{a}{``Width'' is width of window for boxcar and triangle
smoothings, $\sigma$ for Gaussian smoothing, and number of points in window
for variable-width smoothings.}
\end{deluxetable}

\begin{deluxetable}{lrrrr}
\tablewidth{0pt}
\scriptsize
\tablecaption{Flux Density Ratios from $\chi^2$ Minimization
\label{tab_fratios}}
\tablehead{
\colhead{Smoothing} 
 & \colhead{Smoothing}
 & \colhead{}
 & \colhead{}
 & \colhead{}
 \\
\colhead{Function}
 & \colhead{``Width''}
 & \colhead{$S_A / S_B$}
 & \colhead{$S_C / S_B$}
 & \colhead{$S_D / S_B$}
}
\startdata
Boxcar         &  5      & 2.0387 & 1.0368 & 0.3506  \\ 
               & 10      & 2.0383 & 1.0378 & 0.3508  \\ 
               & 15      & 2.0411 & 1.0379 & 0.3508  \\ 
Triangle       &  5      & 2.0387 & 1.0368 & 0.3505  \\ 
               & 10      & 2.0400 & 1.0376 & 0.3507  \\ 
               & 15      & 2.0409 & 1.0379 & 0.3511  \\ 
Gaussian       &  3      & 2.0407 & 1.0378 & 0.3511  \\ 
               &  5      & 2.0411 & 1.0379 & 0.3511  \\ 
               &  7      & 2.0410 & 1.0379 & 0.3508  \\ 
Variable-width &  3      & 2.0406 & 1.0383 & 0.3511  \\ 
boxcar         &  5      & 2.0415 & 1.0384 & 0.3512  \\ 
               &  7      & 2.0413 & 1.0382 & 0.3512  \\ 
Variable-width &  3      & 2.0407 & 1.0383 & 0.3511  \\ 
triangle       &  5      & 2.0414 & 1.0384 & 0.3512  \\ 
               &  7      & 2.0393 & 1.0382 & 0.3512  \\ 
Interpolation  & \nodata & 2.0394 & 1.0374 & 0.3506  \\ 
\enddata
\end{deluxetable}

\subsubsection{Cross-correlation\label{xcorr}}

For the cross-correlation calculations the component B light curve is
once again taken as the control curve.  Before the cross-correlations
are performed, both the comparison and control curves are smoothed and
then divided by their mean values over the observations.  The
normalization by the mean value puts the fractional variations in the
light curves at the same level.
The cross-correlation functions are computed in the time domain, with
the value at each lag calculated in the standard fashion:
\begin{equation}
r_{jk} \equiv \frac{s_{jk}^2}{s_js_k} 
\end{equation}
where $s^2_{jk}$ is the covariance defined by
\begin{equation}
s_{jk}^2 = \frac
{\frac{1}{N-1} 
   \sum_i \left[ \frac{1}{\sigma_{ij}^2+\sigma_{ik}^2}
    (x_{ij}-\bar{x}_j)(x_{ik}-\bar{x}_k)\right]}
{\frac{1}{N} \sum_i \frac{1}{\sigma_{ij}^2+\sigma_{ik}^2}}
\end{equation}
and $N$ is the number of overlapping points (e.g., \cite{6_bevington}).
Note that in $x_{ij}$ or $\sigma_{ij}$, the first subscript refers to a 
particular observation, and the second subscript refers to the
name of the variable under discussion.  Thus $x_{ij}$ is the $i^{th}$
observation of variable $x_j$.
The variance,  $s_j^2$, is given by
\begin{equation}
s_j^2= \frac{
  \frac{1}{N-1} \sum_i \left[\frac{1}{\sigma_i^2}
     (x_{ij}-\bar{x}_j)^2\right]}
  {\frac{1}{N}\sum_i \frac{1}{\sigma_i^2}}
\end{equation}
where the means are weighted means,
\begin{equation}
\bar{x}_j = \frac{\sum_i \left[\frac{1}{\sigma_i^2}x_{ij}\right ]}
{\sum_i \frac{1}{\sigma_i^2}}.
\end{equation}

\begin{figure}
\plotone{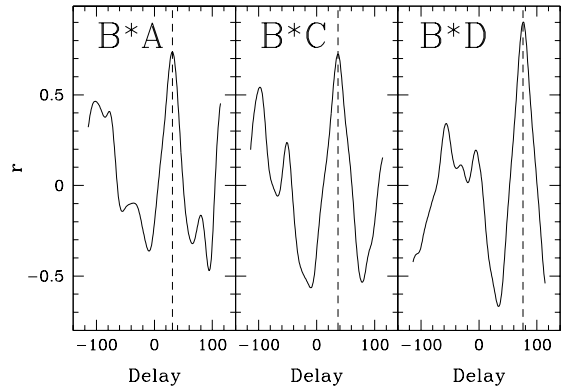}
\caption
{\label{fig_pcorr3} Plot of correlation coefficient vs. lag from
comparison of the light curves of components B and A (left), B and C
(middle), and B and D (right).  Each light curve was smoothed with a
Gaussian with $\sigma = 5$d before the cross-correlations were
calculated.  The maxima (vertical dashed lines) are at lags of 31, 37,
and 76~d.}
\end{figure}

The cross-correlation calculations are repeated for all of the
smoothing functions described in \S\ref{smooth}.  In all cases, clear
peaks in the cross-correlation curves are seen.  Typical curves are
shown in Fig.~\ref{fig_pcorr3}.  The lags at
which the peak correlation coefficients occur are given in
Table~\ref{tab_delays}.  The average displacement between the
cross-correlation and $\chi^2$ minimization lags for each pair of
curves is less than one day.

\subsection{Dispersion Analysis\label{dispersion}}

\begin{figure}
\plotone{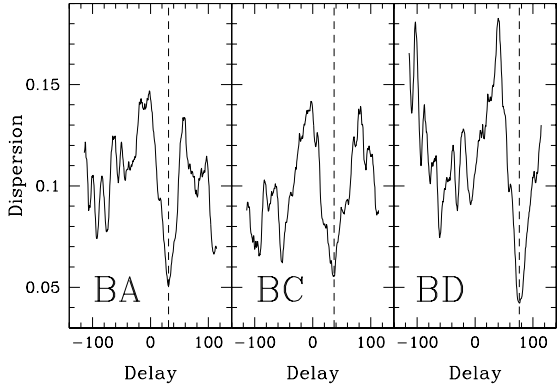}
\caption
{\label{fig_pdisp3} Plot of dispersion spectra from comparison of the
light curves of components B and A (left), B and C (middle), and B and
D (right).  The spectra were calculated with the $D_{4,2}^2$ method
with $\delta = 5.5$d.  The minima (vertical dashed lines) are at lags
of 31, 37, and 77~d.}
\end{figure}

The dispersion analysis methods presented by Pelt et al.\ (1994, 1996,
1998) do not involve any interpolation of the component light curves.
These methods thus have the advantage of avoiding effects introduced
by the interpolation and smoothing associated with the methods
discussed in
\S\ref{smooth}.  The dispersion analysis begins with the construction
of a composite curve $C_k (t_k)$ from two input light curves $A_i$ and
$B_j$.  Curve $A_i$ is not modified, while curve $B_j$ is scaled by a
flux density ratio $\mu$ and shifted in time by a delay $\tau$, i.e.
\begin{equation}
C_k (t_k) = \left\{ \begin{array}{ll}
                    A_i,     & {\rm if}\ t_k = t_i \\
                    \mu B_j, & {\rm if}\ t_k = t_j + \tau
                    \end{array}
            \right .
\end{equation}
(\cite{pelt96}).  We have used a grid of $(\mu, \tau)$ pairs to
construct the $C_k$ curves.  Aside from changing the spacing on the
delay axis to 0.5~d, the grid limits and spacings are the same as
those used in the $\chi^2$\ minimization analysis presented in
\S\ref{chi2}.  The internal dispersion in each curve is calculated,
and the grid point associated with the minimum dispersion is recorded.
In our analysis of the B1608+656 light curves we calculate dispersions
using the non-parametric $D_2^2$ and the one-parameter $D_{4,2}^2$
statistics, where the notation is taken from Pelt et al.\ (1996; note
that these statistics are called $D_1^2$ and $D_2^2$, respectively, in
Pelt et al.\ 1998).  The $D_2^2$ dispersion is calculated using only
immediately adjacent points in the composite curve, with the caveat
that a pair of points only contributes to the dispersion if the two
points are from different input curves.  The $D_{4,2}^2$ dispersion is
similar, but uses all pairs of points that lie within a time range
$\delta$ of each other.  For a detailed description of these
estimates, see Pelt et al.\ (1996).  Table~\ref{tab_disp} gives the
results for the $D_2^2$ analysis and the $D_{4,2}^2$ analysis with
several values of the $\delta$ parameter.  The results do not depend
strongly on the value of $\delta$ and are consistent within the errors
(\S\ref{mc_delay}) with the results from the $\chi^2$\ minimization
and cross-correlation analyses.  The dispersion spectra plotted in
Fig.~\ref{fig_pdisp3} are cuts through the $(\mu, \tau)$ grid at a
value of $\mu$ corresponding to the minimum dispersion.

\begin{deluxetable}{crcccrrr}
\tablewidth{0pt}
\scriptsize
\tablecaption{Dispersion Analysis Results
\label{tab_disp}}
\tablehead{
\colhead{Statistic}
 & \colhead{$\delta$}
 & \colhead{$\Delta t_{BA}$}
 & \colhead{$\Delta t_{BC}$}
 & \colhead{$\Delta t_{BD}$}
 & \colhead{$S_A / S_B$}
 & \colhead{$S_C / S_B$}
 & \colhead{$S_D / S_B$}
}
\startdata
$D_2^2$     & \nodata &  30.5 &  2.0388 &  36.5 & 1.0372 & 74.5 & 0.3505 \\
$D_{4,2}^2$ &  3.5    &  32.0 &  2.0429 &  35.5 & 1.0361 & 74.5 & 0.3501 \\
$D_{4,2}^2$ &  4.5    &  31.5 &  2.0429 &  36.5 & 1.0372 & 74.5 & 0.3502 \\
$D_{4,2}^2$ &  5.5    &  31.0 &  2.0429 &  37.0 & 1.0372 & 76.5 & 0.3509 \\
$D_{4,2}^2$ &  6.5    &  31.0 &  2.0429 &  34.5 & 1.0361 & 76.5 & 0.3509 \\
$D_{4,2}^2$ &  7.5    &  31.5 &  2.0429 &  35.0 & 1.0372 & 76.5 & 0.3509 \\
$D_{4,2}^2$ &  8.5    &  32.0 &  2.0429 &  35.0 & 1.0372 & 77.0 & 0.3512 \\
$D_{4,2}^2$ &  9.5    &  31.5 &  2.0449 &  34.5 & 1.0372 & 77.5 & 0.3512 \\
$D_{4,2}^2$ & 10.5    &  31.5 &  2.0449 &  35.0 & 1.0372 & 77.0 & 0.3512 \\
$D_{4,2}^2$ & 11.5    &  32.0 &  2.0449 &  35.5 & 1.0372 & 77.5 & 0.3512 \\
$D_{4,2}^2$ & 12.5    &  32.0 &  2.0449 &  36.5 & 1.0372 & 78.5 & 0.3512 \\
$D_{4,2}^2$ & 13.5    &  32.0 &  2.0449 &  35.5 & 1.0372 & 78.5 & 0.3512 \\
$D_{4,2}^2$ & 14.5    &  32.5 &  2.0449 &  35.0 & 1.0372 & 78.5 & 0.3512 \\
$D_{4,2}^2$ & 15.5    &  33.0 &  2.0449 &  34.5 & 1.0372 & 77.5 & 0.3512 \\
$D_{4,2}^2$ & 16.5    &  32.5 &  2.0449 &  35.0 & 1.0372 & 79.5 & 0.3516 \\
$D_{4,2}^2$ & 17.5    &  32.5 &  2.0449 &  35.0 & 1.0372 & 79.0 & 0.3512 \\
$D_{4,2}^2$ & 18.5    &  32.0 &  2.0429 &  35.0 & 1.0372 & 79.5 & 0.3512 \\
$D_{4,2}^2$ & 19.5    &  32.0 &  2.0429 &  35.0 & 1.0372 & 79.5 & 0.3512 \\
$D_{4,2}^2$ & 20.5    &  32.0 &  2.0429 &  35.0 & 1.0372 & 80.0 & 0.3516 \\
\enddata
\end{deluxetable}

\subsection{\label{finaldelays}Time Delays}

\begin{figure}
\plotone{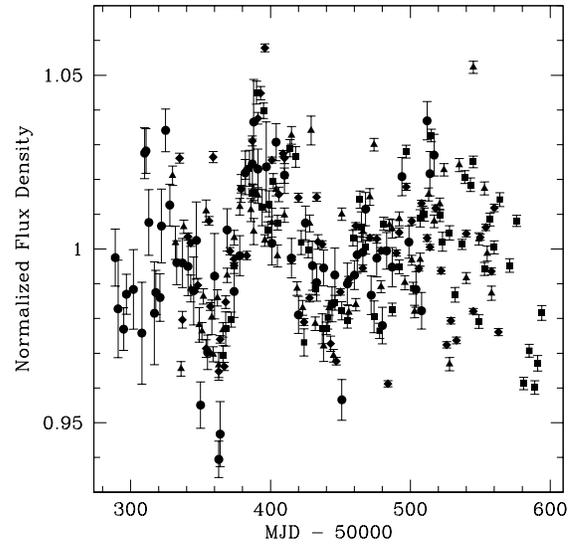}
\caption
{\label{fig_compos} Composite light curve constructed by
normalizing the component light curves and shifting by the lags given
in \S\ref{finaldelays}.  The light curves are represented by squares
(component A), diamonds (B), triangles (C), and circles (D).  
For clarity of presentation, error bars represent only the additive 
component contributed by the RMS noise in the maps (see \S\ref{flat}).}
\end{figure}

\begin{figure}
\plotone{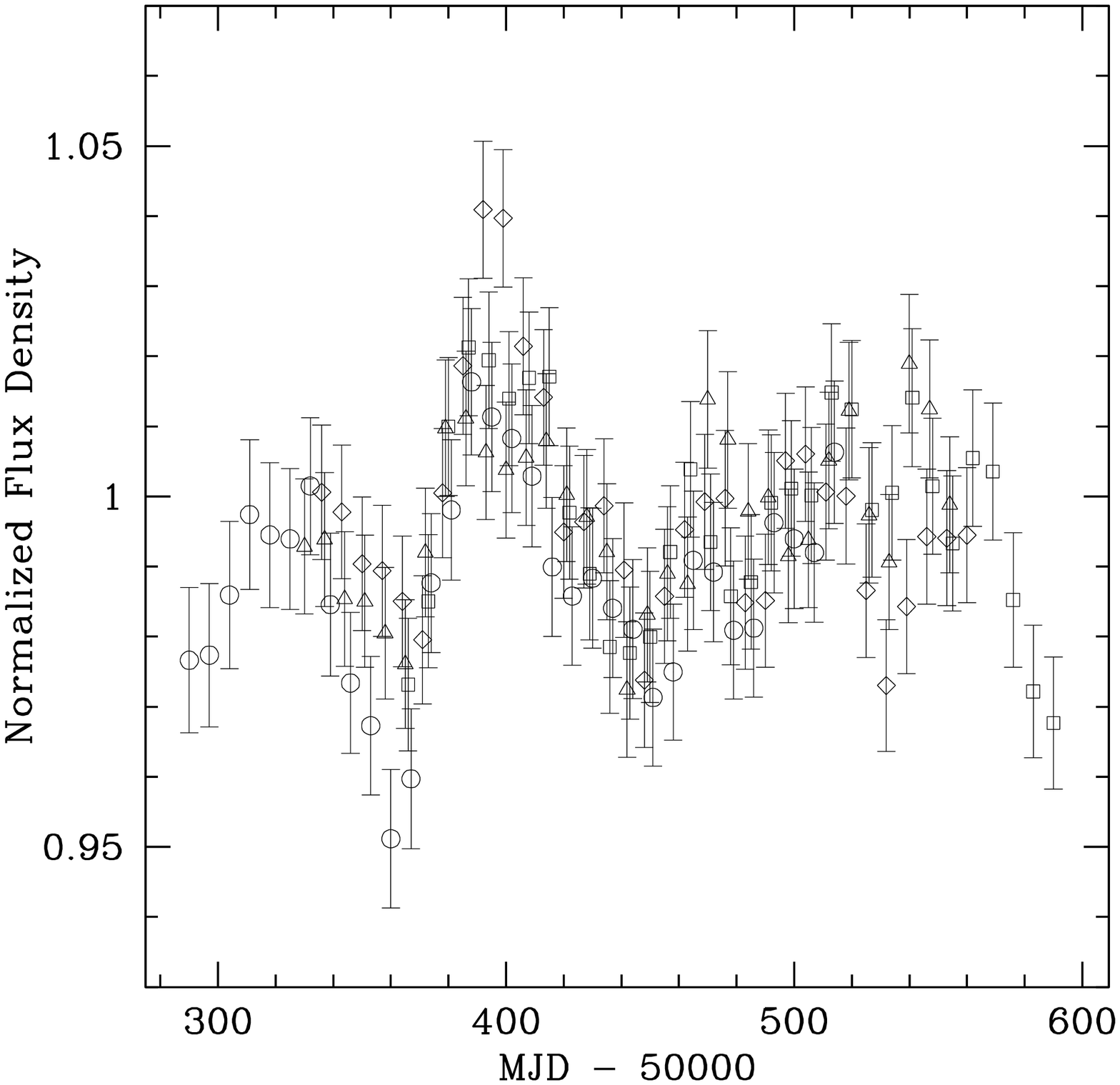}
\caption
{\label{fig_compos_smo} Composite light curve constructed as
in Fig.~\ref{fig_compos}, but using smoothed rather than raw light
curves.  The curves are smoothed with a boxcar of width 15~d.  
For clarity, only one out of every 7 points is shown for each curve.
}
\end{figure}

There is some scatter in the results presented in
Tables~\ref{tab_delays} and \ref{tab_disp}, which is not surprising
considering the low levels of variation and sparse sampling of the
light curves.  However, the scatter is small compared with both the
length of the time delays and the uncertainties in the delays that we
find from the Monte Carlo simulations in \S\ref{mc_delay}.  For each
of the three methods used to find the delays ($\chi^2$ minimization,
cross-correlation, and dispersion analysis), we take the median values
of the delays in Tables~\ref{tab_delays} and \ref{tab_disp}.  The
three median values for each delay are then averaged to obtain $\Delta
t_{BA}$ = 31~d, $\Delta t_{BC}$ = 36~d, and $\Delta t_{BD}$ = 76~d.
The median flux density ratios, computed in a similar manner, are
found to be $S_A / S_B = 2.0418$, $S_C / S_B = 1.0376$, and $S_D / S_B
= 0.3512.$ We shift the light curves by the mean delays and normalize
them using the mean flux density ratios to create a composite light
curve of the background source (Fig.~\ref{fig_compos}).  A composite
curve constructed from the smoothed and interpolated component light
curves is shown in Fig.~\ref{fig_compos_smo}.

\section{Monte Carlo Simulations}

\subsection{Significance of Light Curve Correlations}

The variations seen in the B1608+656 light curves are not large,
either in a fractional {\em or} absolute sense, compared to what has
been seen in other lens systems.  The fractional variations seen in
B0218\+357, 0957+561, and PG~1115+080 are all two to three times larger
than those seen in B1608+656 (\cite{6_b0218}; Kundi\'{c} et al.\
1995,1997; \cite{6_s1115}), and PKS~1830$-$211 shows 50\% variations
in flux density (although the time delay measurement is based on
smaller variations; \cite{6_l1830}).  A sceptic might argue that the
correlations between the B1608+656 light curves are not significant
and could be duplicated by any set of light curves containing random
scatter about a constant value.

In theory, the value of the correlation coefficient can be used to
assess the significance of the correlation.  With the coefficient,
$r$ calculated as described in \S\ref{xcorr}, the probability of 
obtaining a value $\geq r$ from two {\em uncorrelated} curves of 
Gaussian-distributed random variables is
\begin{equation}
P_c(r,N)=2 \int_{|r|}^1  P_r(\rho,\nu) d\rho
\end{equation}
where
\begin{equation}
P_r(r,\nu) = \frac{1}{\sqrt{\pi}} 
\frac{\Gamma[(\nu+1)/2]}{\Gamma(\nu/2)}\;(1-r^2)^{(\nu-2)/2}
\label{eqn_bev_prob}
\end{equation}
and $\nu = N - 2$ (e.g., \cite{6_bevington}).  However, the smoothing
and interpolation performed on the sparsely sampled B1608+656 light
curves makes the interpretation of the significance of the value of
$r$ complicated.  The difficulty lies in assessing the number of {\em
independent} points in the curves at each lag.  Monte Carlo
simulations show that the number of independent points cannot be
estimated simply as the width of the overlap region divided by the
width of the smoothing window (i.e., the number of smoothing windows
in the overlap region).  This calculation {\em underestimates} the
number of independent points in the region.  Because it is difficult
to determine the significance of the correlation analytically, we
perform Monte Carlo simulations to find the significance empirically.
In the simulations, we calculate correlations between light curves
consisting of randomly distributed data.  To create random light
curves with the same distribution of flux densities as seen in the
data, we simply randomize the time series for the component light
curves while preserving the flux densities at their measured values.
By randomizing the times at which the flux densities are measured, we
destroy any possible correlations between the curves.  Each light
curve is randomized independently to avoid correlations at zero lag
which are associated with measurement errors.

\begin{figure}
\plotone{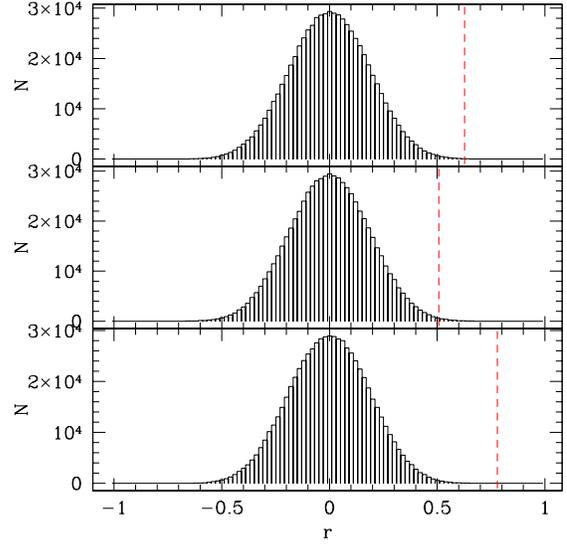}
\caption
{\label{fig_randplot} Distribution of cross-correlation
values obtained from 3000 Monte Carlo simulations of randomized light
curves.  The distributions presented in this figure are obtained by
smoothing the light curves with a boxcar of width 10~d.  
The dashed vertical lines represent the peak cross-correlation
values obtained from the real data.}
\end{figure}

The simulations are conducted with 3000 sets of randomized curves.
Each set of light curves is processed in the manner described in
\S\ref{xcorr} and produces three sets of correlation curves (B--A, B--C,
and B--D).  All values of the correlation coefficient are recorded.
The distributions of the cross-correlation values obtained from the
10~d boxcar smoothing scheme are shown in Fig.~\ref{fig_randplot}.
The empirical probabilities of obtaining at least the observed peak
values, which are indicated by the vertical dashed lines in the
figure, from uncorrelated curves are all low, with $P(|r| \geq |r_A|)
= 2.8 \times 10^{-4}$, $P(|r| \geq |r_C|) = 4.3 \times 10^{-3}$, and
$P(|r| \geq |r_D|) < 1.5 \times 10^{-6}$.  It is even more unlikely
that three pairs of randomized curves could produce three such
anomalously high cross-correlation peaks.  There can thus be no
significant doubt that the correlations we measure are real.

\subsection{\label{mc_delay}Uncertainties in Time Delays and Flux Density 
Ratios}

The time delay measurement uncertainties contribute directly to the
error budget for measuring $H_0$ with a gravitational lens.  In
particular, the fractional uncertainties in the time delays contribute
a matching fractional uncertainty in $H_0$, i.e.,  
\begin{equation}
\frac{(\sigma_H)_{delay}}{H_0} = \frac{\sigma_{\Delta t}}{\Delta t}.
\label{eqn_H0_err}
\end{equation} 
We estimate the uncertainties in the time delays by performing Monte
Carlo simulations of the observations.  In the simulations we assign
time delays between the pairs of light curves and then see how well we
can recover the input delays.  We also use the simulations to estimate
the uncertainties in the flux density ratios, which are necessary for
modeling the lensing potential (Paper 2).

We produce fake curves with the same characteristics as our real data
by smoothing the composite light curve (Fig.~\ref{fig_compos}) with a
10~d boxcar filter.  This normalized and smoothed curve is the master
light curve for the simulations, representing the assumed true
behavior of the background source.  The offsets between the points of
the composite curve and the master curve are distributed as a
zero-mean Gaussian with $\sigma = 0.014$.  Note that this is a {\em
fractional} value since all of the light curves have been normalized
to create the master light curve.  The Gaussian distribution is used
to generate the random offsets for the simulations.  The appropriate
rescaling of the random offsets is achieved through
Equation~\ref{eqn_mc}.  We note that smoothing procedure used in
constructing the master curve destroys information on possible
short-timescale variations. Thus, if such short-term variations do
exist, their effects will be attributed to measurement and calibration
errors in this method.  However, given the low level of variability of
the background source during the observations and the sparse sampling,
it is difficult to distinguish any possible short-term variations from
measurement error.  Thus, we accept $\sigma = 0.014$ as an indication
of the measurement error, but with the caveat that the uncertainties
in the time delays derived from the Monte Carlo simulations described
below may be over-estimated.

For each round of the simulation we generate four sparsely-sampled
fake light curves.  The flux density for component $j$ at a time $t_i$
in the fake curves is given by:
\begin{equation}
S_j(t_i) = S_B\,R_j ( S_0(t_i + \Delta t_j)  + n_{ij} ), 
\quad j = {A,B,C,D}
\label{eqn_mc}
\end{equation}
where $S_0(t)$ is the normalized master flux density, the $R_j$ are the input
flux density ratios (2.0418, 1.000, 1.0376, and 0.3512), the $\Delta
t_j$ are the input time delays (31~d, 0~d, 36~d, and 76~d), and
$n_{ij}$ is the random offset.  
The four curves are sampled with the pattern used in the
observations (see Table~\ref{tab_obs}).  The flux density error at
each point in the sparsely-sampled curves is set to the observed flux
density error for that epoch (see \S\ref{flat}).

\begin{figure}
\plotone{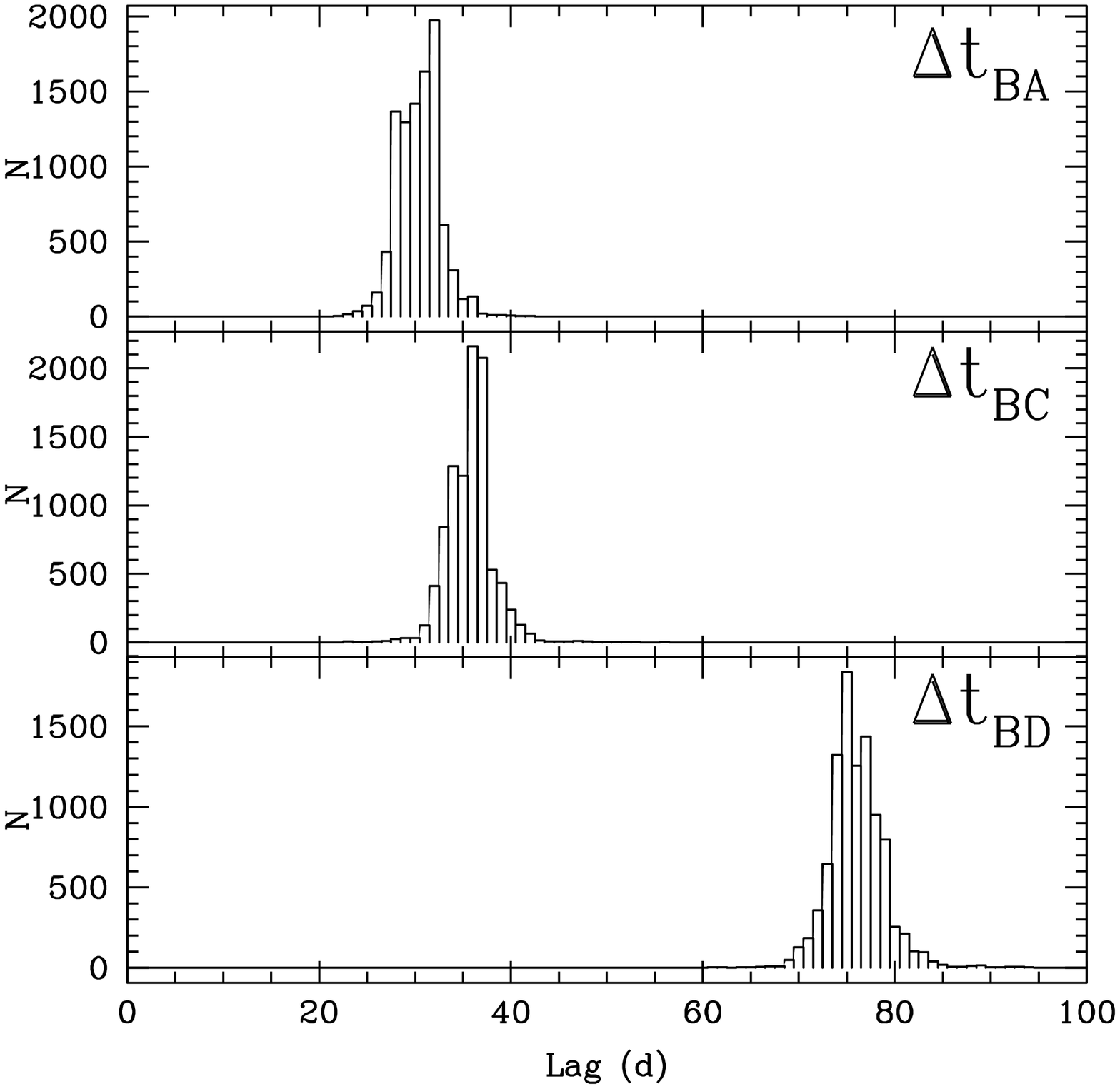}
\caption
{Distribution of time delays recovered from cross-correlation analysis of
10,000 Monte Carlo simulations of the B1608+656 component light curves.  
The distributions provide estimates of the uncertainties in the time delays.
\label{fig_mc_xc} }
\end{figure}

The fake curves are processed in the manner described in
\S\ref{timedelay}.  The best-fit time delays and flux density ratios
for each simulation are recorded.  Histograms of typical distributions
of time delays from 10,000 repetitions of the above procedure are
shown in Fig.~\ref{fig_mc_xc}.  The distributions are non-Gaussian,
both in the shape of the peak of the distribution and in the long tail
of outliers at negative delays.  Thus, we determine the confidence
limits by finding the range of delays inside which 95\% of the
simulation results lie, rather than fitting a Gaussian to the
distribution.  The limits are chosen such that the minimum range of
delays that encloses both the median value and 95\% of the simulation
results is found.  The results for the $\chi^2$ minimization and
cross-correlation techniques are given in Table~\ref{tab_mc}.  We
conservatively take the broader distribution in each case as our
estimate of the 95\% confidence contours.  Thus, we estimate the time
delays to be $\Delta t_{BA} = 31\pm 7$~d, $\Delta t_{BC} = 36\pm 7$~d,
and $\Delta t_{BD} = 76^{+9}_{-10}$~d at 95\% confidence.  

We have also run Monte Carlo simulations in which the sampling pattern
is varied, following the method used in Biggs et al. (1999).  The
distributions of time delays do not differ significantly from those
presented above.

\begin{deluxetable}{cccccc}
\tablewidth{0pt}
\scriptsize
\tablecaption{Monte Carlo Simulation Results
\label{tab_mc}}
\tablehead{
 &
 &
 & \multicolumn{3}{c}{Confidence Intervals}
\\
\colhead{Quantity}
 & \colhead{Input Value}
 & \colhead{Method\tablenotemark{a}}
 & \colhead{68\%}
 & \colhead{90\%}
 & \colhead{95\%}
}
\startdata
$\Delta t_{BA}$ & 31     & 1 & 28.6 -- 33.0 & 26.7 -- 34.9 & 24.4 -- 37.3 \\
$\Delta t_{BC}$ & 36     & 1 & 33.5 -- 38.7 & 31.8 -- 40.3 & 29.0 -- 43.3 \\
$\Delta t_{BD}$ & 76     & 1 & 73.2 -- 78.9 & 71.1 -- 81.0 & 67.4 -- 84.7 \\
$\Delta t_{BA}$ & 31     & 2 &  29  --  34  &  27  --  36  &  25  --  38  \\
$\Delta t_{BC}$ & 36     & 2 &  33  --  38  &  31  --  40  &  29  --  42  \\
$\Delta t_{BD}$ & 76     & 2 &  71  --  77  &  68  --  80  &  66  --  82  \\
$S_A / S_B$     & 2.0418 & 1 & 2.0386 -- 2.0506 & 2.0344 -- 2.0549 & 2.0322 -- 2.0570 \\
$S_C / S_B$     & 1.0376 & 1 & 1.0340 -- 1.0401 & 1.0318 -- 1.0422 & 1.0306 -- 1.0433 \\
$S_D / S_B$     & 0.3512 & 1 & 0.3500 -- 0.3525 & 0.3492 -- 0.3534 & 0.3487 -- 0.3539 \\
\enddata
\tablenotetext{a}{Method used to calculate delays or flux density ratios.  1 = $\chi^2$ minimization.  2 = cross-correlation.}
\end{deluxetable}

\section{Discussion}

The goal of monitoring a gravitational lens system is to measure time
delays that can then be combined with a model of the lensing potential
to produce a measurement of $H_0$.  We have been successful in
measuring the three independent time delays in the B1608+656 system.
Paper 2 presents a model for the B1608+656 system.  The model is based
on the time delays and flux density ratios presented here, and on
positions from VLBA (\cite{6_16082045vlba}) and HST observations of
the system (\cite{6_hstlens}).  The lensing
potential contains contributions from the two lensing objects seen in
HST images of the system (\cite{6_hstlens}), each of which is modeled
as an elliptical isothermal mass distribution.  The effects of varying
the positions of the lensing galaxies, of changing the nature of the
lensing galaxy cores (singular or non-singular), and of departing from
an isothermal profile are all explored.  The best-fit model is
obtained through a simulated annealing process.  It correctly
reproduces the positions and flux density ratios of the lensed images
(with the exception of the D/A flux density ratio).  Most importantly,
the predicted time delay {\em ratios} match the observed time delay ratios
to within 1\%.  Although the individual time delays depend on the
Hubble constant ($\Delta t_i \propto h^{-1}$), the time delay ratios
have no $H_0$ dependence.  Thus, the model for any lens system with
more than two images {\em must} correctly reproduce the observed time
delay ratios if it is to be used in the determination of $H_0$.  In
this sense, gravitational lenses which produce more than two images
can put stronger constraints on lens models than can two image lenses,
as long as the time delays can be measured.

\begin{figure}
\plotone{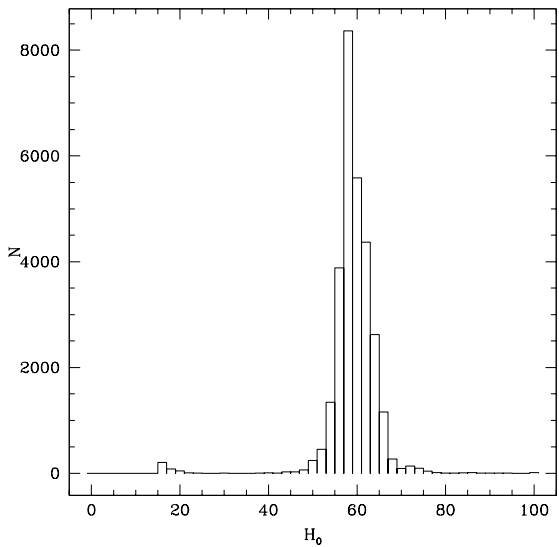}
\caption
{Distribution of estimates of $H_0$ from Monte Carlo simulations.
This distribution is formed by converting the time delay simulation
results into three $H_0$ distributions, and then combining the
individual distributions.\label{fig_hall} }
\end{figure}

The B1608+656 system is the first four-image lens for which the three
independent time delays have been measured and a model correctly
reproduces the time delay ratios.  The best-fit isothermal model from
Paper 2 predicts time delays of $\Delta t_{BA} = 18.0~h^{-1}$~d,
$\Delta t_{BC} = 21.4~h^{-1}$~d, and $\Delta t_{BD} = 44.9~h^{-1}$~d.
Combining these predicted values with the observed time delays gives
three individual determinations of $H_0$: $(H_0)_{BA} = 58.1$,
$(H_0)_{BC} = 59.4$, and $(H_0)_{BD} = 59.1~{\rm km}\,{\rm
s}^{-1}\,{\rm Mpc}^{-1}$.  We combine these by calculating a
weighted mean of $(H_0)_{1608} = 59.0~{\rm km}\,{\rm s}^{-1}\,{\rm
Mpc}^{-1}$, where the weights are derived from the uncertainties in
the time delays from \S\ref{mc_delay}.  Note that the B-D value
dominates the mean since it has the smallest fractional uncertainty.
In order to estimate the uncertainty on $(H_0)_{1608}$, we use the
results of the Monte-Carlo simulation presented in \S\ref{mc_delay}.
The distribution of time delays from the simulations are converted
into distributions of $H_0$ by dividing the predicted delays from the
Paper 2 model by the simulation results.  The resulting $H_0$
distributions from the B-A, B-C, and B-D delays are combined into an
overall distribution, which is shown in Fig.~\ref{fig_hall}.  The
confidence limits are estimated by finding the ranges that enclose 68,
90, and 95\% of the data.  The resulting 95\% confidence limits are
$(H_0)_{1608} = 59^{+8}_{-7}~{\rm km}\,{\rm s}^{-1}\,{\rm
Mpc}^{-1}$.  These confidence limits are very close to the statistical
95\% limits derived from the lens model, which includes $H_0$ as a
model parameter ($59^{+7}_{-6}~{\rm km}\,{\rm s}^{-1}\,{\rm
Mpc}^{-1}$; Paper 2).

The above estimates of the uncertainties in the determination of $H_0$
have not included the systematic effects from the choice of the radial
mass profile in the lens modeling.  The estimated systematic error is
$\pm15{\rm km}~{\rm s}^{-1}\,{\rm Mpc}^{-1}$ (Paper 2).  It may be
possible to reduce this error by modeling the extended lensed stellar
emission from the background source, as has been done with the radio
Einstein ring MG~1654+1346 (\cite{6_csk1654}).  This modeling approach
is being conducted by Surpi \& Blandford (private communication).

\section{Summary}

We have presented the results of an intensive program of monitoring
the four-image lens system B1608+656 with the VLA.  The component light
curves show $\sim$5\% variations in flux density from which we have
measured the three independent time delays in this system: $\Delta
t_{BA} = 31\pm 7$~d, $\Delta t_{BC} = 36\pm 7$~d, and $\Delta t_{BD} =
76^{+9}_{-10}$~d.  These time delays are combined with the mass model
of the lens presented in Paper 2 to give $H_0 = 59^{+8}_{-7}~{\rm
km}\,{\rm s}^{-1}\,{\rm Mpc}^{-1}$ at 95\% confidence (statistical)
$\pm 15~{\rm km}\,{\rm s}^{-1}\,{\rm Mpc}^{-1}$ (systematic).  The
statistical uncertainties represent the 95\% confidence interval.  The
statistical uncertainties in the time delays can be reduced if a
stronger variation in the background source is observed, while the
systematic uncertainties may be reduced through the inclusion of the
lensed extended emission in the lens modeling process. Our previous
observations have shown that the background source in this system has
varied by as much as 15\% in the past, so we are conducting another
program of monitoring.  If a stronger variation is detected in the new
data, the uncertainties on the time delays will be reduced and the
accuracy of the measurement of $H_0$ with this system will be
improved.

\acknowledgments 

For generously donating portions of their VLA observing time to help
us fill in gaps in our time coverage, we are grateful to Erik Leitch,
Brian Mason, Jackie Hewitt, Cathy Trotter, Gillian Knapp, Michael
Rupen, and James Gunn.  The observations would not have been possible
without the expertise and help provided by the VLA analysts and
operators.  We thank the anonymous referee for helpful comments.  For
useful discussions, we are indebted to Lori Lubin, Andy Biggs, Roger
Blandford, Geoff Bower, Ketan Desai, Debbie Haarsma, Phillip Helbig,
Jackie Hewitt, Tomislav Kundi\'c, Erik Leitch, Chung-pei Ma, Mark
Metzger, Chris Moore, Gerry Neugebauer, Frazer Owen, Michael Rupen,
David Rusin, Martin Shepherd, and Ed Turner. This work is supported in
part by the NSF under grant \#AST 9420018 and by the European
Commission, TMR Program, Research Network Contract ERBFMRXCT96-0034
``CERES.''


\begin{thebibliography}{}

\bibitem[Bevington 1969]{6_bevington}Bevington, P.\ R. 1969, {\em
Data Reduction and Error Analysis for the Physical Sciences},
(New York: McGraw-Hill)

\bibitem[Biggs et al.\ 1999]{6_b0218}Biggs, A.\ D., Browne, I.\ W.\ A.,
Helbig, P., Koopmans, L.\ V.\ E., Wilkinson, P.\ N., \& Perley, R.\ A.
1999, \mnras, 304, 349

\bibitem[Blandford \& Narayan 1986]{6_bn86}Blandford, R.\ \& Narayan, R.,
1986, 310, 568

\bibitem[Blandford \& Narayan 1992]{6_bn}Blandford, R.\ D.\ \& Narayan, R.
1992, \araa, 30, 311

\bibitem[Browne et al.\ 1998]{6_jvas2}Browne, I.\ W.\ A., Patnaik, A.\ R.,
Wilkinson, P.\ N., \& Wrobel, J.\ M.\ 1998, \mnras, 293, 257

\bibitem[Fassnacht et al.\ 1996]{6_f1608}Fassnacht, C.\ D., Womble D.\ S.,
Neugebauer, G., Browne, I.\ W.\ A., Readhead, A.\ C.\ S., Matthews, K.,
\& Pearson, T.\ J. 1996, \apjl, 460, L103

\bibitem[Fassnacht et al.\ 1999]{6_16082045vlba}Fassnacht, C.\ D. et al.
1999, in preparation.

\bibitem[Haarsma et al.\ 1999]{6_dh0957}Haarsma, D.\ B., Hewitt, J.\ N.,
Leh\'{a}r, J., \& Burke, B.\ F. 1999, \apj, 510, 64

\bibitem[H\"ogbom 1974]{6_clean}H\"ogbom, J. 1974, \apjs, 15, 417

\bibitem[Impey et al.\ 1998]{6_i1115}Impey, C.\ D., Falco, E.\ E.,
Kochanek, C.\ S., Leh\'ar, J., McLeod, B.\ A., Rix, H.-W., Peng, C. Y.,
\& Keeton, C.\ R. 1998, \apj, 509, 551

\bibitem[Jackson et al.\ 1997]{6_hstlens}Jackson, N.\ J., Nair, S.,
Browne, I.\ W.\ A.\ 1997, in Observational Cosmology with the New
Radio Surveys, eds., M.\ Bremer, N.\ Jackson \& I.\ Perez-Fournon,
(Dordrecht: Kluwer) 315

\bibitem[Kochanek 1995]{6_csk1654}Kochanek, C.\ S. 1995, \apj, 445, 559

\bibitem[Koopmans \& Fassnacht 1999]{6_lkmodel}Koopmans, L.\ V.\ E. \&
Fassnacht, C.\ D. 1999, \apj, submitted (Paper 2)

\bibitem[Kundi\'c et al.\ 1995]{6_k0957_1}Kundi\'c, T. et al.\ 1995, \apjl,
455, L5

\bibitem[Kundi\'c et al.\ 1997]{6_k0957_2}Kundi\'c, T. et al.\ 1997, \apj,
482, 75

\bibitem[Lovell et al.\ 1998]{6_l1830}Lovell, J.\ E.\ J., Jauncey, D.\ L.,
Reynolds, J.\ E., Wieringa, M.\ H., King, E.\ A., Tzioumis, A.\ K.,
McCulloch, P.\ M., \& Edwards, P.\ G. 1998, \apjl, 508, L51

\bibitem[Moore \& Hewitt 1997]{6_m0414}Moore, C.\ B.\ \& Hewitt, J.\ N. 1997,
\apj, 491, 451

\bibitem[Myers et al.\ 1995]{6_m1608}Myers, S.\ T., et al.\ 1995, \apjl, 447, L5

\bibitem[Myers et al.\ 1999]{6_class}Myers, S.\ T., et al.\ 1999, in preparation

\bibitem[Oscoz et al.\ 1997]{6_o0957}Oscoz, A., Mediavilla, E., Goicoechea,
L.\ J., Serra-Ricart, M. \& Buitrago, J. 1997, \apjl, 479, L89

\bibitem[Patnaik et al.\ 1992]{6_jvas1}Patnaik, A.\ R., Browne, I.\ W.\ A.,
Wilkinson, P.\ N., \& Wrobel, J.\ M. 1992b, \mnras, 254, 655

\bibitem[Pelt et al.\ 1994]{pelt94}Pelt, J., Hoff, W., Kayser, R., Refsdal, S.,
\& Schramm, T. 1994, \aap, 286, 775

\bibitem[Pelt et al.\ 1996]{pelt96}Pelt, J., Kayser, R., Refsdal, S.,
\& Schramm, T. 1996, \aap, 305, 97

\bibitem[Pelt et al.\ 1998]{pelt98}Pelt, J., Hjorth, J., Refsdal, S.,
Schild, R. \& Stabell, R. 1998, \aap, 337, 681

\bibitem[Perley \& Taylor 1999]{6_vlacalib}Perley, R.\ A.\ \& Taylor, G.\ B.
1999, {\em VLA Calibrator Manual}

\bibitem[Refsdal 1964]{6_refsdal}Refsdal, S. 1964, \mnras, 128, 307

\bibitem[Schechter et al. 1997]{6_s1115}Schechter, P.\ L., et al.\ 1997, \apjl,
475, L85

\bibitem[Schneider, Ehlers, \& Falco 1992]{6_gravlenses}Schneider, P., Ehlers,
J., \& Falco, E.\ E. 1992 {\em Gravitational Lenses}, (New York: 
Springer-Verlag)

\bibitem[Shepherd 1997]{6_difmap}Shepherd, M.\ C. 1997, in 
{\em Astronomical Data Analysis Software and Systems VI}, eds. 
G.\ Hunt \& H.\ E.\ Payne, (ASP Conference Series, v125) 77

\bibitem[Snellen et al.\ 1995]{6_s1608}Snellen, I.\ A.\ G., de Bruyn,
A.\ G., Schilizzi, R.\ T., Miley, G.\ K. \& Myers, S.\ T. 1995, \apjl, 447, L9


\bibitem[Walsh et al.\ 1979]{6_0957discovery}Walsh, D., Carswell, R.\ F., \& 
Weymann, R.\ J. 1979, {\em Nature}, 279, 381

\bibitem[Weymann et al.\ 1980]{6_1115discovery}Weymann, R.\ J., Latham,
D., Roger, J., Angel, P., Green, R.\ F., Liebert, J.\ W., Turnshek,
D.\ A., Turnshek, D.\ E., \&  Tyson, J.\ A. 1980, {\em Nature}, 285, 641

\bibitem[White \& Peterson (1994)]{6_xcorr_reverb}White, R.\ J.\ \& Peterson,
B.\ M. 1994, \pasp, 106, 879

\bibitem[Wilkinson et al.\ 1998]{6_jvas3}Wilkinson, P.\ N., Browne, I.\ W.\ A.,
Patnaik, A.\ R., Wrobel, J.\ M., \& Soratia, B. 1998, \mnras, 300, 790

\end{thebibliography}
\end{document}